# A Unified Intracellular pH Landscape with SITE-pHorin: a Quantum-Entanglement-Enhanced pH Probe

Shu-Ang Li[1,4], Xiao-Yan Meng[1,2,4], Su Zhang[3,4], Ying-Jie Zhang[1,2], Run-Zhou Yang[1], Dian-Dian Wang[1,2], Yang Yang[1], Pei-Pei Liu[1], Jian-Sheng Kang[1,2,5]*

**Affiliations:**

[1] Clinical Systems Biology Laboratories, The First Affiliated Hospital of Zhengzhou University; Zhengzhou 450052, China.

[2] The Academy of Medical Sciences, Zhengzhou University; Zhengzhou 450001, China.

[3] National Clinical Research Center for Infectious Diseases, Guangdong Provincial Clinical Research Center for Tuberculosis, Shenzhen Third People's Hospital, Southern University of Science and Technology; Shenzhen, 518112, China.

[4] These authors contributed equally

[5] Lead contact

*Corresponding author: kjs@zzu.edu.cn (J.S.K.)

## Abstract

An accurate map of intracellular organelle pH is crucial for comprehending cellular metabolism and organellar functions. However, a unified intracellular pH spectrum using a single probe is still lack. Here, we developed a novel quantum entanglement-enhanced pH-sensitive probe called SITE-pHorin, which featured a wide pH-sensitive range and ratiometric quantitative measurement capabilities. Subsequently, we measured the pH of various organelles and their sub-compartments, including mitochondrial sub-spaces, Golgi stacks, endoplasmic reticulum, lysosomes, peroxisomes, and endosomes in COS-7 cells. For the long-standing debate on mitochondrial compartments pH, we measured the pH of mitochondrial cristae as $6.60 \pm 0.40$, the pH of mitochondrial intermembrane space as $6.95 \pm 0.30$, and two populations of mitochondrial matrix pH at approximately $7.20 \pm 0.27$ and $7.50 \pm 0.16$, respectively. Notably, the lysosome pH exhibited a single, narrow Gaussian distribution centered at 4.79 ± 0.17. Furthermore, quantum chemistry computations revealed that both the deprotonation of the residue Y182 and the discrete curvature of deformed benzene ring in chromophore are both necessary for the quantum entanglement mechanism of SITE-pHorin. Intriguingly, our findings reveal an accurate pH gradient (0.6-0.9 pH unit) between mitochondrial cristae and matrix, suggesting prior knowledge about ΔpH (0.4-0.6) and mitochondrial proton motive force (pmf) are underestimated.

**Keywords**: pH probe; mitochondria; quantum entanglement; SITE-pHorin; unified intracellular pH map; Golgi stacks; lysosome; endoplasmic reticulum; peroxisome; mitochondrial cristae; mitochondrial intermembrane space; mitochondrial matrix

**Introduction**

The precise regulation of pH levels is essential for optimal metabolic health and overall functioning. Physiological blood pH is tightly restrained to a narrow range from pH 7.35 to pH 7.45, so that acidemia refers to pH less than 7.35, while alkalemia is related to pH greater than 7.45 (Burger and Schaller, 2023). Deviations from the norm can significantly impact the occurrence and progression of diverse diseases (Freeman et al., 2023). Specifically, aberrant pH variations within cellular organelles, such as mitochondria, have been closely linked to serious ailments such as cancer(Vyas et al., 2016) and neurodegenerative diseases(Devine and Kittler, 2018). Similarly, alterations in the pH of the Golgi apparatus can result in abnormal glycosylation, which has been linked to the development of cancers and cutis laxa formation(Rivinoja et al., 2012). Dysregulation of pH within lysosomes can cause lysosome dysfunction, potentially contributing to various diseases including neurodegenerative diseases(Wang et al., 2018), inflammation and autoimmune diseases(Ge et al., 2015), lipid and glucose metabolism disorders(Gu et al., 2021), among others. Hence, meticulous monitoring of pH fluctuations within organelles can enhance our understanding of complex disease mechanisms, offering valuable insights for diagnosis and therapeutic interventions.

Various strategies have been employed to detect organelle pH, including pH microelectrodes(Sommer et al., 2000), capillary electrophoresis(Chen and Arriaga, 2006), Nuclear Magnetic Resonance (NMR)(Kulichikhin et al., 2007), targeted avidin and fluorescein-biotin(Wu et al., 2000), fluorescent dye(Ramshesh and Lemasters, 2018) and fluorescent protein(Llopis et al., 1998), (Jankowski et al., 2001), (Porcelli et al., 2005) , among others. Different organelle pH levels have been monitored using various fluorescent and dyes and proteins, as summarized in Table S1. However, these methods exhibit inconsistencies and various limitations. For instance, pH microelectrodes require cell manipulation, which can damage cells and cannot simultaneously monitor multiple organelles in real-time. NMR measurements require expensive equipment and specialized technical expertise, rendering them unsuitable for live experiments. Given these limitations, a single pH-sensitive fluorescent protein might offer a more consistent and preferable solution. pH-sensitive fluorescent proteins provide advantages such as minimal cell damage, heightened sensitivity, and enhanced specificity when paired with organelle-targeted markers.

A variety of pH-sensitive fluorescent protein probes have been developed, including pHluorins(Jankowski et al., 2001), (Reifenrath and Boles, 2018), (Liu et al., 2017), (Mahon, 2011), T-Sapphire(Zapata-Hommer and Griesbeck, 2003), deGFP4(Hanson et al., 2002a), mNectarine(Johnson et al., 2009), pHRed(Tantama et al., 2011), SypHer(Poburko et al., 2011), pHTomato(Li and Tsien, 2012), pHuji(Shen et al., 2014), pHoran(Shen et al., 2014), pH-Lemon(Burgstaller et al., 2019), pHLARE(Webb et al., 2021), pHmScarlet(Liu et al., 2021a), mOrange2(Sarker et al., 2022), and others. However, these fluorescent proteins still possess certain limitations. For example, some exhibit lower quantum yield, limiting their fluorescence signal strength. Others have a narrow pH response range, restricting their utility in measuring pH across a wide spectrum. Additionally, some of these proteins feature single excitation and single emission characteristic, which may not be ideal for measuring pH in all types of organelles.

In our study, we have successfully engineered a quantum entanglement-enhanced pH-sensitive and ratiometric fluorescent protein called SITE-pHorin (a _si_ngle excitation and _t_wo _e_missions _pH_ sens_or_ prote_in_). This protein features one major excitation wavelength and two distinct emission wavelengths, enabling precise measurements of organelle pH. SITE-pHorin exhibits a higher quantum yield (QY), ensuring enhanced fluorescence signal. Additionally, it demonstrates a wider pH response range, spanning from pH 3.5 to 9.0. The organelle pH is determined by calculating the fluorescent intensity ratio between the two emission wavelengths, providing accurate and quantitative pH measurements. Particularly, our findings reveal a more accurate estimation of the pH gradient (0.6-0.9 pH) across mitochondrial inner membrane, suggesting that the prior knowledge about ΔpH (0.4-0.6) has been underestimated. Furthermore, quantum chemistry computations reveal that the deprotonation of the residue Y182 and the discrete curvature of deformed benzene ring in chromophore are both necessary for the quantum entanglement mechanism of SITE-pHorin.

## Results

### The spectral characteristics of mTurquoise2, mTurquoise2 S65T and W66Y mutants

In light of its faster maturation, high photostability, longer mono-exponential lifetime and the highest quantum yield up to 93%, mTurquiose2, a monomeric cyan fluorescent protein, was selected as the foundation for developing a bright and pH-sensitive fluorescent protein(Goedhart et al., 2012). However, with a very low pKa of 3.1, mTurquiose2 did not exhibit significant pH sensitivity in high pH buffers, limiting its utility in detecting organellar pH levels. Previous studies have shown that the GFP mutants S65T and Y66H improved pH sensitivity(Kneen et al., 1998). Therefore, we hypothesized that the two amino acids S65 and Y66 in mTurquiose2 might also influence its pH sensitivity. To test this hypothesis, we conducted mutagenesis at residues 65 and 66 of the mTurquiose2 chromophore. As anticipated, the two mTurquiose2 mutants (S65T and W66Y) displayed enhanced pH sensitivity compared to mTurquiose2 (Fig. 1).

To validate their pH sensitivity and spectral characteristics, we expressed and purified mTurquiose2, mTurquiose2 S65T and mTurquiose2 W66Y protein using *E.coli* BL21(DE3) strains. The excitation and emission spectra of mTurquiose2, mTurquiose2 S65T, and

mTurquiose2 W66Y were recorded using 96-well microplate fluorometer, in buffers ranging from pH 3.5 to pH 9.0. As anticipated, the S65T mutant did not alter the spectral characteristics of mTurquiose2. Both mTurquiose2 wild type (Fig. 1a, 1b) and S65T (Fig. 1c, 1d) had a maximal excitation at 440nm and a maximal emission wavelength at 475 nm, respectively. Interestingly, the S65T mutation enhanced the pH responsiveness of mTurquoise2, similar to the effect observed in GFP S65T(Elsliger et al., 1999). In contrast, the W66Y mutation restored the chromophore of mTurquoise2 to that of GFP (S65-Y66-G67), resulting in spectral characteristics resembling the wild-type green fluorescent protein found in the Victoria Bioluminescent jellyfish. It displayed a maximum excitation wavelength of 395 nm and a maximum emission wavelength of 510 nm (Fig. 1e, 1f). Intriguingly, the W66Y mutation also increased the pH sensitivity of mTurquoise2 within the pH range of 3.5 to 9.0 compared to mTurquoise2. By normalizing the fluorescence intensity of the emission peak at 475 nm (mTurquiose2), 475 nm (mTurquiose2 S65T) and 510 nm (mTurquiose2 W66Y), we confirmed that both mTurquiose2 mutants, S65T and W66Y, exhibited greater sensitivity (Fig. 1h). Additionally, the full width at half maximum (FWHM) of excitation and emission, as well as the pH value for maximum emission, were also depicted (Fig. 1g), further complementing the characterization of mTurquoise2, mTurquoise2 S65T, and mTurquoise2 W66Y.

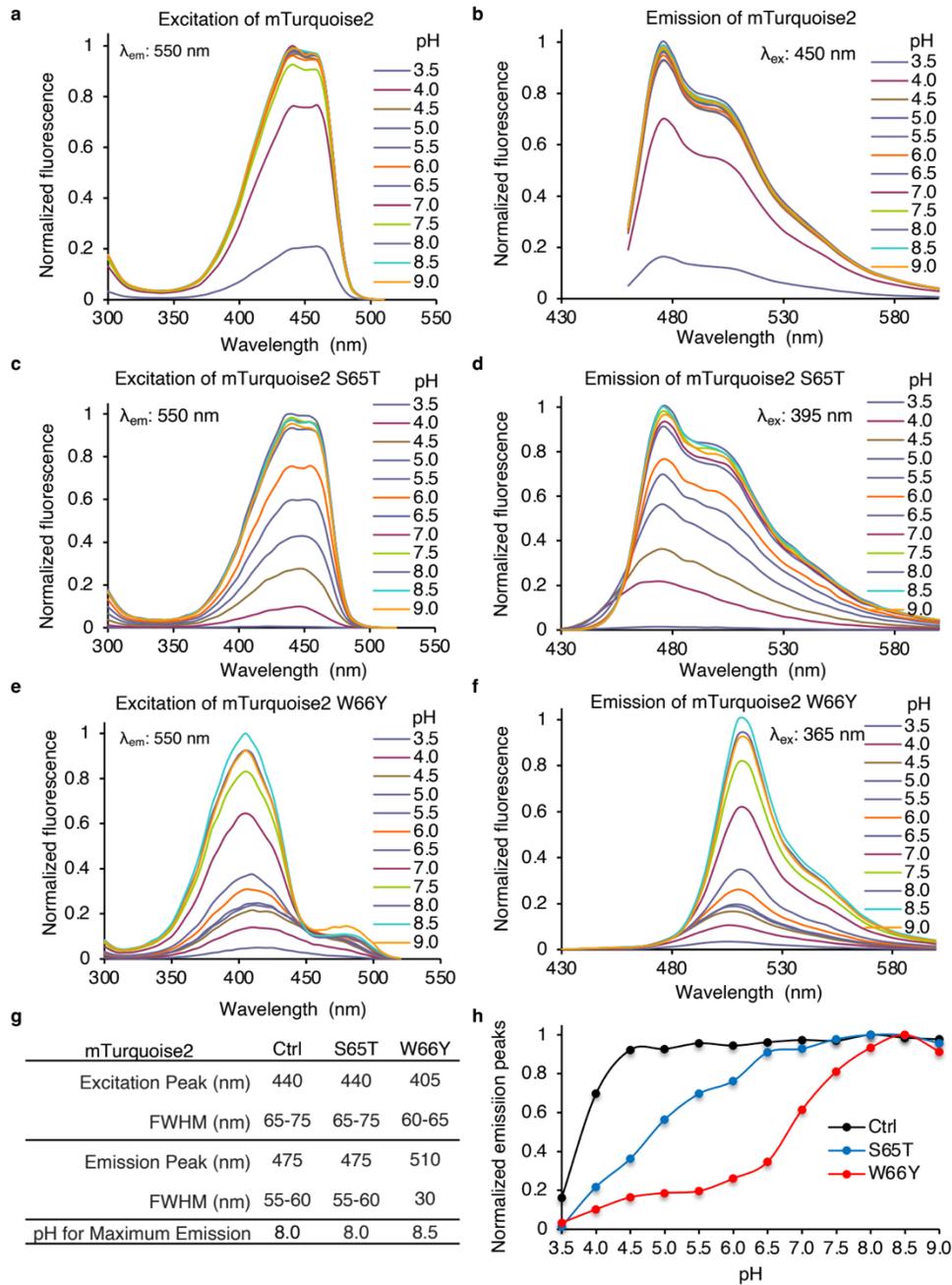

**Fig. 1| The spectral characteristics of mTurquoise2, mTurquoise2 S65T and W66Y mutants.**

**The key residue D148 of mTurquoise2 and its saturation mutations**

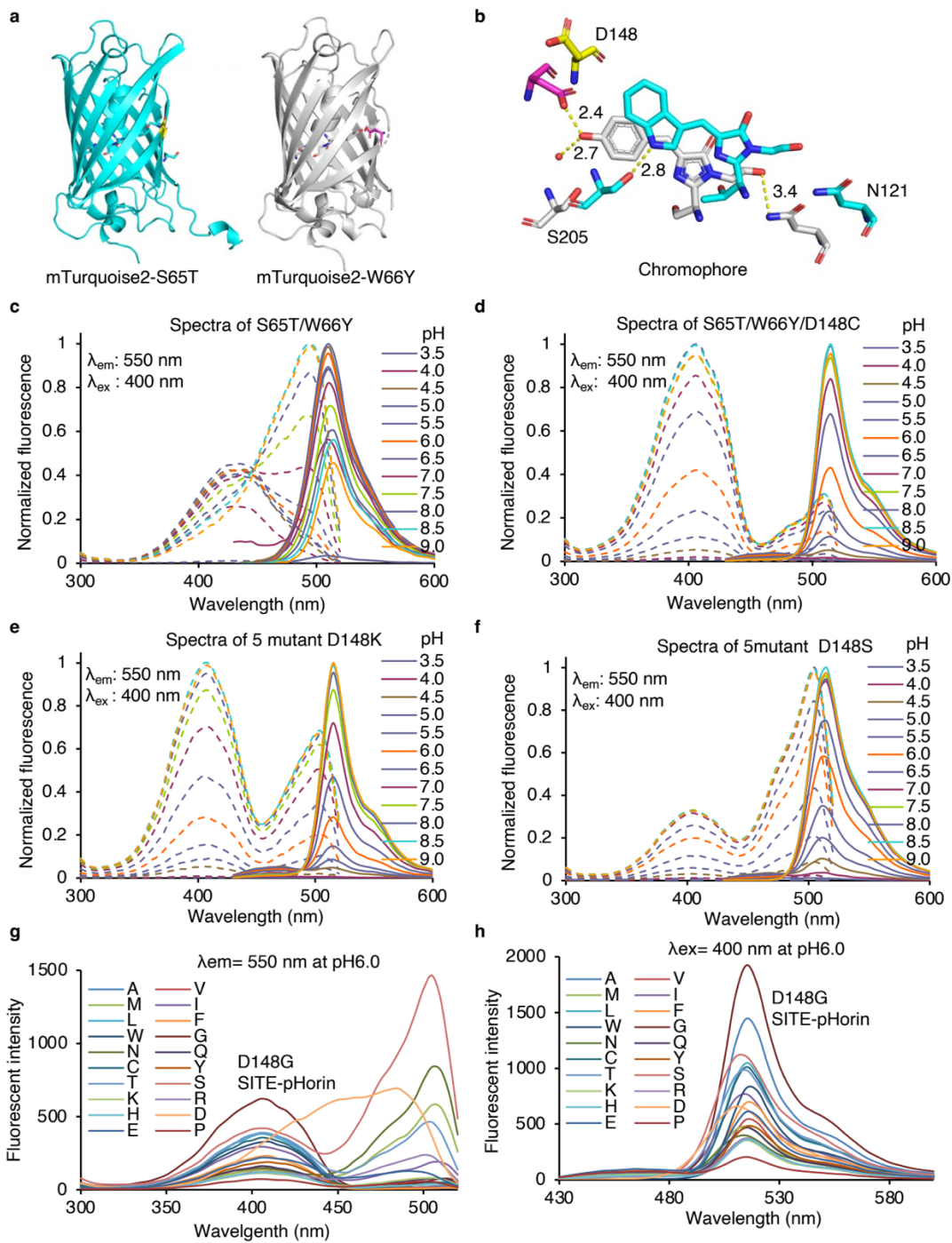

**Fig. 2| The key residue D148 of mTurquoise2 and its saturation mutations.**

Although mTurquiose2 S65T and W66Y were more pH-sensitive than mTurquiose2, they were still not ideal as the optimal pH-sensitive probe. An optimal probe should possess one major excitation and two emissions characteristics, enabling ratiometric measurement for accurate pH monitoring across all organelles. Ratiometric measurement offers clear advantages, reducing or eliminating distortions in experimental data caused by factors like probe abundance, illumination stability, and imaging settings. To further improve the spectra characteristic of mTurquoise2, we aimed to elucidate the pH-sensitive mechanism by examining the crystal structures of the mTurquiose2 S65T and W66Y mutants (Fig. 2a). Crystals of mTurquiose2 S65T and W66Y were successfully obtained under crystallization condition at pH 6.5, as described in the methods section. Their structures were determined using molecular replacement and refined at resolutions of 2.00 Å and 2.10 Å, respectively (Table S2). In mTurquiose2 W66Y, D148 was in a loop conformation, whereas in mTurquiose2 S65T, it formed a β-sheet (Fig. 2a). A comparative analysis of the crystal structures of mTurquiose2 S65T with W66Y suggested that the chromophore (S65-Y66-G67, dyed in white) of mTurquiose2 W66Y formed hydrogen bonds with D148 and N121, while the chromophore (T65-W66-G67, dyed in cyan) of mTurquiose2 S65T formed a hydrogen bond with S205 (Fig. 2b). Subsequently, we analyzed the channels of mTurquoise2, mTurquoise2 S65T and mTurquoise2 W66Y using the software MOLE2.5. Interestingly, mTurquoise2 had no channel, but mTurquoise2 S65T and mTurquoise2 W66Y exhibited a channel allowing water molecules to pass through easily (Fig. S2). This suggest that mTurquiose2 S65T and W66Y might be more sensitive to the change of hydrogen ion concentration than mTurquoise2. Based on the proposed hydrogen bonds network of mTurquiose2 and mTurquiose2 S65T, it is evident that they can form hydrogen formed bonds with V61, Q94, R96, and S205 residues (Fig. S3a, S3c). In contrast, mTurquiose2 W66Y formed additional hydrogen bonds with the E222 residue (Fig. S4a). Electrostatic surfaces analysis revealed that the carbonyl moiety of the glycine residue in mTurquiose2 chromophore (SWG) was converted to an oxhydryl moiety and became deprotonated, carrying a negative charge (Fig. S3b). However, the carbonyl moiety of mTurquoise2 S65T chromophore (TWG) had a neutral charge (Fig. S3d). In contrast, the mTurquoise2 W66Y chromophore (SYG) not only contained the oxhydryl moiety of glycine, which could be deprotonated similarly to the mTurquoise2 chromophore, but also featured the phenolic hydroxyl moiety of tyrosine as a site for deprotonation. Thus, mTurquiose2 W66Y possessed two moieties capable of bonding and dissociating hydrogen ions. In conclusion, biochemical assays and analysis of crystal structures confirmed the enhanced pH-sensitivity of mTurquoise2 W66Y and mTurquoise2 S65T compared to mTurquoise2 in the physiologic pH range.

Compared with deGFP4(Hanson et al., 2002a), a reference for the characteristic of single excitation and two emissions, we also examined whether the C48 and T203 residues could influence the pH sensitivity of the protein. To investigate this, we introduced mutations in the C48 and T203 residues, examined the excitation and emission spectra of the single mutant C48S and T203C (Fig. S1), and found that these two mutants had no effects on the pH sensitivity and fluorescent spectra of mTurquoise2 (Fig. 1a, 1b). Subsequently, we analyzed the excitation and emission spectra of the double mutant S65T/W66Y (Fig. 2c), which exhibited the characteristic of a dual excitation and single emission pattern. In addition, we examined various double mutants and triple mutants (Fig. S5) involving the combinations of the S65T, W66Y, D148C and T203C mutations. We found that the double mutant (W66Y/D148C) (Fig. S5c), the triple mutants (S65T/W66Y/D148C) (Fig. 2d) and (W66Y/D148C/T203C) (Fig. S5g) exhibited favorable characteristics of one major excitation peak and two emission peaks.

Considering the above results and the imperfect spatial structure of D148 in the mTurquoise2 W66Y mutant (Fig. 2a), we speculated that D148 might be a key residue for pH sensitivity and fluorescent spectra. Consequently, we created saturation mutagenesis of D148 based on the mTurquiose2 quadruple mutant (C48S/S65T/W66Y/T203C) (Fig. S5h). Among these mutants with five mutations of mTurquoise2, D148K (Fig. 2e) and D148S (Fig. 2f) exhibited the characteristic of two excitations and two emissions. The D148K mutant displayed a major excitation peak at 400 nm, and the spectra of D148E, D148Q, D148I, D148H, D148W, D148P and D148Y mutants were similar (Fig. S6). On the other hand, D148S had a main excitation peak around at 500 nm, and D148R, D148N, D148M and D148T were close analogs (Fig. S7). Ideally, we discovered some mutants, namely D148G, D148A, D148F, D148C, D148L and D148V (Fig. S8), which exhibit the characteristic of one major excitation and two emissions. Among these mutants, we observed that the D148G mutant displayed the highest excitation efficiency in a pH 6.0 buffer solution (Fig. 2g) and an emission intensity (Fig. 2h) among twenty D148 mutants. Therefore, we named this mutant with C48S/S65T/W66Y/T203C/D148G mutations as SITE-pHorin. SITE-pHorin possessed the desired attribute of one major excitation and two emissions, making it a promising probe for quantitative analysis. As suggested by the pun of SITE, our goal was to acquire an intracellular pH map for organelles using the SITE-pHorin as a pH probe, making it necessary to further characterize its pH sensitivity.

**The pH sensitive characteristics of SITE-pHorin**

Currently, the pH-sensitive fluorescent protein deGFP4 (Hanson et al., 2002a) is the only available ratiometric dual-emission probe with single excitation, and lacking subsequent applications for pH measurements due to its low quantum yield (QY). We analyzed and compared the emission (Fig. 3a, 3b) and excitation (Fig. 3c, 3d) spectra of SITE-pHorin (Fig. 3a, 3c) and deGFP4 (Fig. 3b, 3d) in the buffers with varied pH from 3.5 to 9.0. Notably, SITE-pHorin exhibited stronger fluorescence intensities (Fig. 3a) and higher excitation efficiencies (Fig. 3c) than the corresponding properties of deGFP4, especially under alkaline conditions. Similarly, both SITE-pHorin and deGFP4 displayed a weak and acid-enhanced emission peak around 465 nm (the insets of Fig. 3a, 3b). Furthermore, we plotted the logarithms of the ratio of fluorescent intensities of the two emission peaks at 515 nm and 465 nm in the pH-varied buffers for SITE-pHorin and deGFP4, and demonstrated that SITE-pHorin exhibited a broader and more linear pH-sensitive response than deGFP4 (Fig. 3e). As a summary, the optical properties of SITE-pHorin and deGFP4 are tabulated in Fig. 3f. SITE-pHorin displayed a three-times higher QY of 0.56 at pH 8.0 compared to the QY 0.15 of deGFP4. Meanwhile, both SITE-pHorin and deGFP4 had low QYs under the condition of pH 5.0. To further validate the pH-sensitive properties of SITE-pHorin at the protein level, we expressed the SITE-pHorin protein using *E.coli* BL21(DE3) and measured its fluorescent intensity in buffers with pH varied from 3.5 to 9.0 (Fig. 3g). The fluorescent ratios of the emission channels (alkaline sensitive > 500 nm and acid sensitive < 480 nm) acquired under pH 3.5-9.0, the ratios were illustrated by pseudo-colors with their scales on the left sides of three panels in Fig. 3g. As demonstrated, the wide range of the scales also vividly represented the broad pH sensitivity of SITE-pHorin.

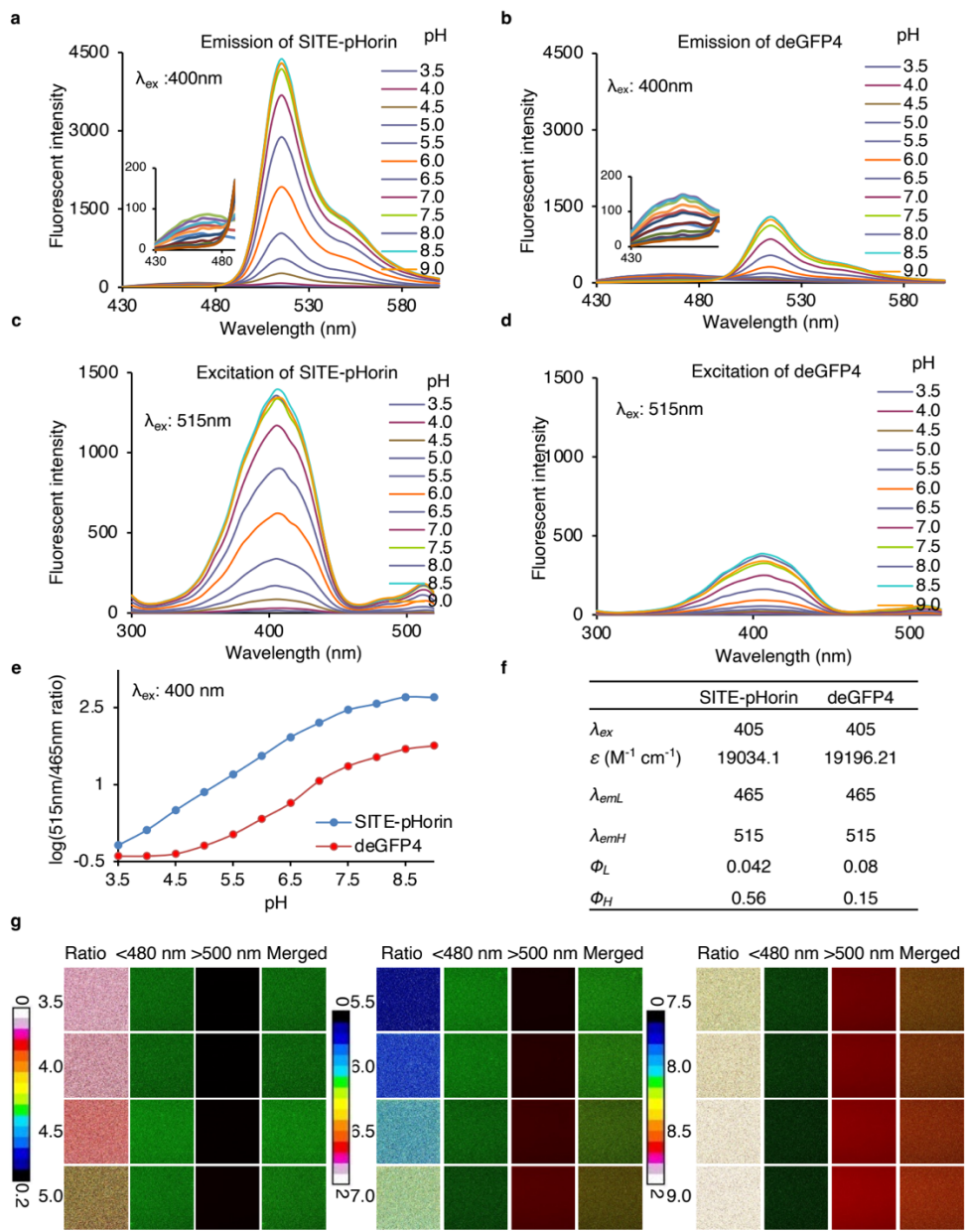

**Fig. 3| The pH sensitive characteristics of SITE-pHorin.**

**The quantum entanglement mechanism of the pH ultra-sensitivity of SITE-pHorin**

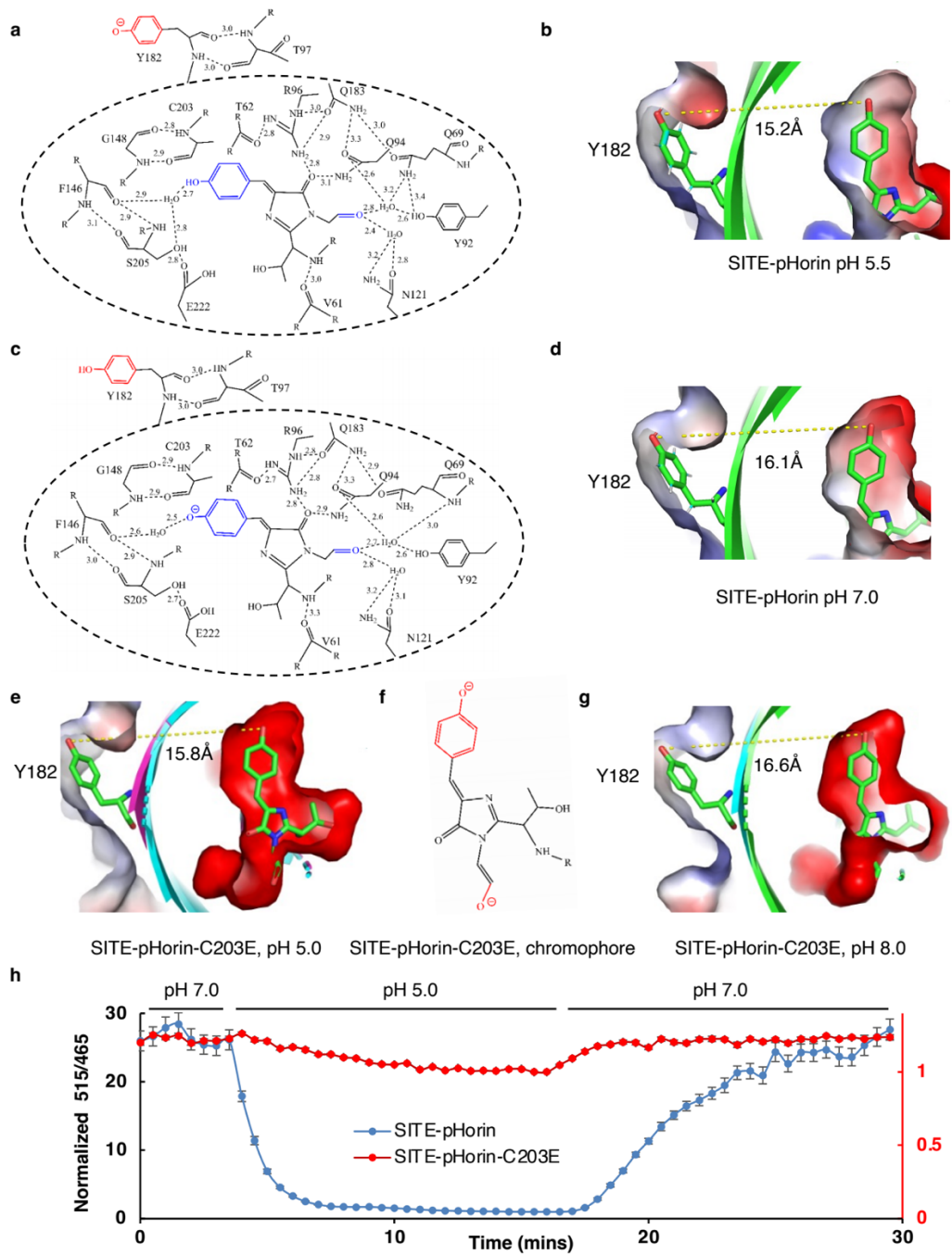

**Fig. 4| The quantum entanglement mechanism of the pH ultra-sensitivity of SITE-pHorin.**

To investigate the pH-sensitive mechanism of SITE-pHorin, we obtained its crystal structure under both high pH 7.0 and low pH 5.5 crystalized conditions. The crystal structures were obtained through molecular replacement and refined to the resolutions of 2.30 Å and 2.183 Å resolution, respectively (Table S2). Analysis of the proposed hydrogen bonds network and electrostatic surfaces of the SITE-pHorin chromophore revealed an intriguing observation. Remarkably, the phenolic hydroxyl moiety of Y182, a residue located outside the β-barrel structure, deprotonated under pH 5.5, which couldn't be explained by the classical theory of chemistry since the pKa of phenol was around 10. Meanwhile, the phenolic hydroxyl moieties of the Y182 residue and the chromophore exhibited an opposite status of protonation or deprotonation under both low pH and high pH conditions (Fig. 4a-d). Such phenomena were absent from the published structures of deGFP1(Hanson et al., 2002b) (Fig. S9) and GFP S65T(Elsliger et al., 1999) (Fig. S10) under different pH conditions. Interestingly, we also observed that the phenolic hydroxyl moiety of Y182 residue in GFP S65T was also deprotonated at low pH (Fig. S10a) but the phenolic hydroxyl moiety of chromophore remained deprotonation at low and high pH conditions (Fig. S10). Meanwhile, the distance between the phenolic hydroxyl moieties of the Y182 and the chromophore in GFP S65T remained constant under different pHs. In contrast, the distances in SITE-pHorin were measured at 15.2 angstroms (Å) at pH 5.5 and 16.1 Å at pH 7.0, respectively. These phenomena led us to speculate that the simultaneous opposite protonation and deprotonation between the chromophore of SITE-pHorin and Y182 could be only explained with quantum entanglement (see more discussion and theoretical modeling in methods), which might lead to a quantum effect ($n = 1/2$) and result in the pH ultra-sensitivity of SITE-pHorin (Fig. S11).

To further confirm the potential quantum mechanism for the pH ultra-sensitivity of SITE-pHorin, we introduced a C203E mutant, which was the residue forming hydrogen bond with G148 (Fig. 4a), to interfere the quantum effect of SITE-pHorin with an extra negative charge.

Subsequently, we analyzed the crystal structure of SITE-pHorin C203E at pH 5.0 and pH 8.0, with resolutions of 1.66 Å and 1.57 Å, respectively (summarized in Table S2). In contrast to the hydrogen bond network and electrostatic surface of SITE-pHorin, we observed that the phenolic hydroxyl moiety of the chromophore in SITE-pHorin_C203E carried two negative charges at pH 5.5 and pH 7.0 (Fig. 4e-g, S12). Consequently, the C203E mutant eliminated the quantum entanglement between phenolic hydroxyl moieties of the Y182 residue and the chromophore in SITE-pHorin. Furthermore, we checked the fluorescent dynamics of SITE-pHorin and its C203E mutant by live imaging and switching mediums with a pH 7.0 or a pH 5.0. In Fig. 4h, the results demonstrated that SITE-pHorin_C203E (red curve and red vertical axis) lost the pH ultra-sensitivity of SITE-pHorin (blue curve). Therefore, the structure and live imaging experiments of the single mutation (C203E) supported that the mechanism of quantum entanglement was necessary for the pH ultra-sensitivity of SITE-pHorin.

**SITE-pHorin targeted to organelles and their sub-compartments**

To accomplish an intracellular pH site map for organelles using the SITE-pHorin, we first evaluated the performance of SITE-pHorin for live cell imaging, and demonstrated that the fluorescent ratios of the emission channels exhibited a wide range of scales and good signal-noise ratios of SITE-pHorin expressed in COS-7 cells under varied pH buffers from 3.5 to 9.0 (Fig. 5a). Next, we targeted SITE-pHorin into the luminal side of various organelles and their sub-compartments. Therefore, through extensive literature review and experimental validations, we identified suitable markers for organelles (Fig. 5b), including Histone H2B as a classic nucleus marker (Musinova et al., 2011), ATP synthase g subunit as a mitochondria cristae marker (identified in the Cryo-EM structure of the mammalian ATP synthase)(Gu et al., 2019), 4Cox8 as a mitochondria matrix marker (consisting of four repeat Cytochrome c oxidase subunit 8)(Kadenbach and Hüttemann, 2015), (Ma and Ding, 2021), (Ross, 2011), Ferrochelatase (Fech) as a mitochondria intermembrane space marker (Nagaraj et al., 2009), (Watanabe et al., 2013), (Sakaino et al., 2009), Dnase2B as a lysosome marker (Ohkouchi et al., 2013), (Shpak et al., 2008), Calreticulin as an endoplasmic reticulum marker (Houen et al., 2021), (Michalak et al., 2009), ST6GAL1 as a trans-Golgi marker (Hassinen et al., 2010), (Khoder-Agha et al., 2019), MGAT2 as a media-Golgi marker (Linders et al., 2022), (Hassinen et al., 2010), GP73 as a cis-Golgi marker (Kladney et al., 2000), (Hu et al., 2011), (Bachert et al., 2007), (Liu et al., 2021b), and SITE-pHorin-SKL as a peroxisome marker (where the last two amino acids YK of SITE-pHorin was mutated to SKL(Miura et al., 1992), (Wolins and Donaldson, 1997), (Elleuche and Pöggeler, 2008). TFR1 was chosen as an endosome marker(Kawabata, 2019) (Fig. 5b). Detailed information about these markers was available in Table S3. The targeting specificities and qualities of specific organelle-targeted probes of SITE-pHorin were validated and confirmed via confocal imaging (Fig. 5c). Interestingly, the fluorescent distribution of the TFR1-tagged SITE-pHorin suggested that TFR1 positive vesicles could be divided into at least two populations with different pHs (Fig. 5c).

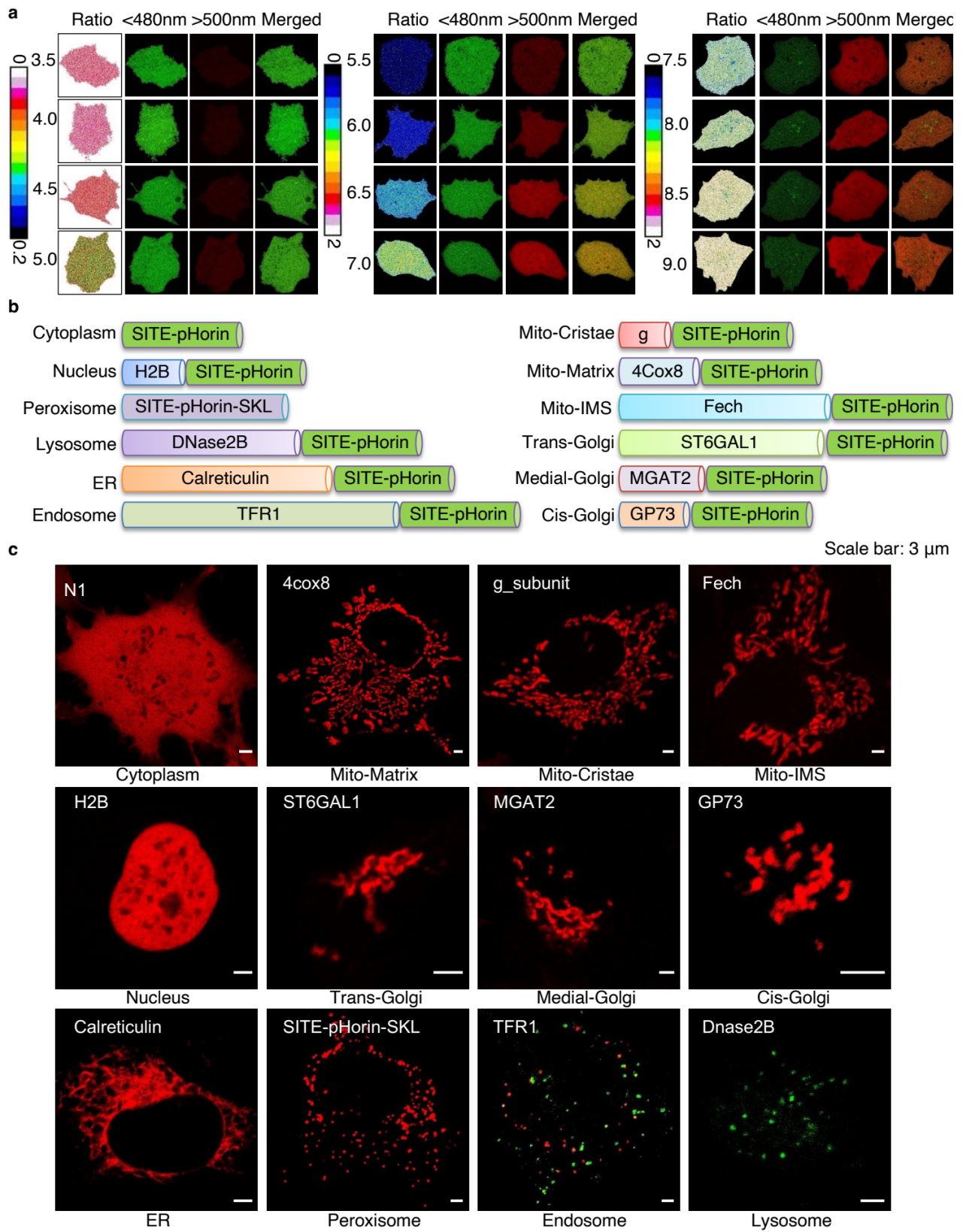

**Fig. 5| SITE-pHorin targeting to intracellular organelles and their sub-compartments.**

## The pH map of intracellular organelles

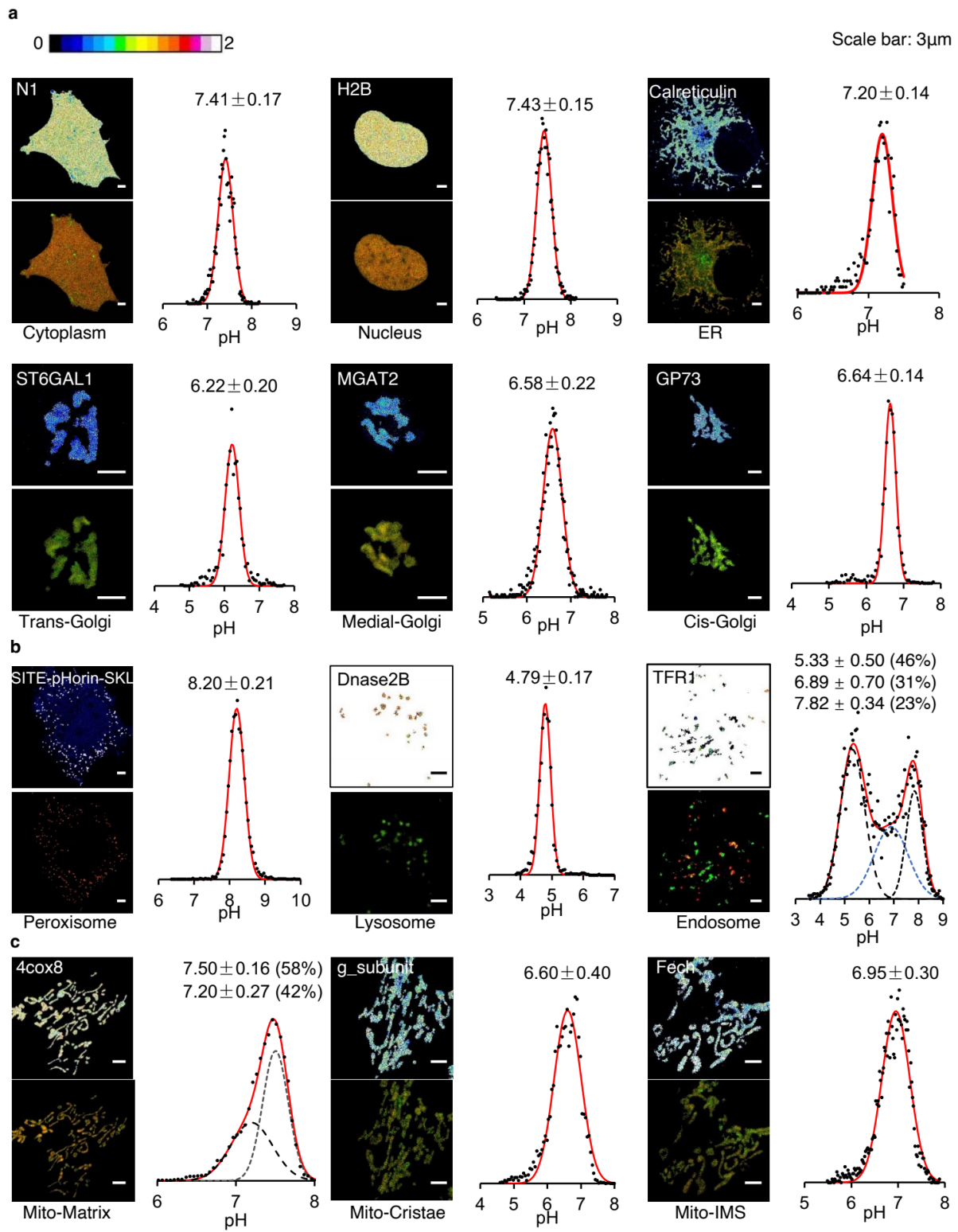

**Fig. 6| The pH and distributions of intracellular organelles.**

Finally, to measure the pH of organelles, we analyzed approximately 100 COS-7 cells transfected with each organelle-targeted SITE-pHorin plasmid. The measured ratios were converted to pH with the equation (7) (Methods). Background noise and autofluorescence were corrected for accurate pH measurement using the equation (9) (Methods), particularly for organelles with strong autofluorescence, such as mitochondrial sub-compartments. Furthermore, to obtain precise pH information for organelles and their sub-compartments, the noise and autofluorescence-corrected pH measurements were further fitted with single or multiple Gaussian distributions (Fig. 6). The cytoplasmic pH was determined to be 7.41 ± 0.17, while the nuclear pH was approximately 7.43 ± 0.15. The endoplasmic reticulum displayed a pH range around 7.20 ± 0.14. As is well known, the pH within the Golgi apparatus exhibited a gradient across its three sub-stacks, with the trans-Golgi pH at 6.22 ± 0.20, the medial-Golgi pH at 6.58 ± 0.22, and the cis-Golgi pH at 6.64 ± 0.14 (Fig. 6a). In addition, the peroxisome was found to have an alkaline pH of 8.20 ± 0.21. Conversely, the lysosome maintained an acidic pH of 4.79 ± 0.17. Intriguingly, our observations revealed the presence of three distinct types of endosomes labelled with TFR1, characterized by pH values of 5.33 ± 0.50 (46%), 6.89 ± 0.70 (31%), and 7.82 ± 0.34 (23%), with the majority being acidic endosomes (Fig. 6b).

Importantly, for mitochondrial sub-compartments (Fig. 6c), we found two populations of mitochondria with different matrix pH values: 7.50 ± 0.16 (58%) and 7.20 ± 0.27 (42%), respectively. Previous reports of mitochondrial matrix of pH values (Table S1) were likely overestimated due to uncorrected mitochondrial autofluorescence (discussed in Methods). The pH of the mitochondrial cristae was determined to be 6.60 ± 0.40, and the pH of mitochondrial intermembrane space was 6.95 ± 0.30. These measurements revealed a pH gradient of approximately 0.6-0.9 pH units between mitochondrial cristae space and mitochondrial matrix. This finding is of great significance for cellular energetics and has been underestimated for a long time in the field of life science, especially in the mitochondrial field. Moreover, the pH value of mitochondrial intermembrane space around 6.95 ± 0.30 corrects the long-standing misconception that the mitochondrial outer membrane is non-specifically permeable to all low-molecular-weight solutes (< 5 KDa via VDAC) (Lemasters, 2007), showing natural gradients of mitochondrial compartments between the intermembrane space and cristae, and between the intermembrane space and cytosol.

As a summary, we have created an illustrative cartoon to summarize the first unified pH map of organelles and their sub-compartments measured using the single pH probe SITE-pHorin, which possesses the rare property of the quantum entanglement-enhanced pH ultra-sensitivity (Fig. 7).

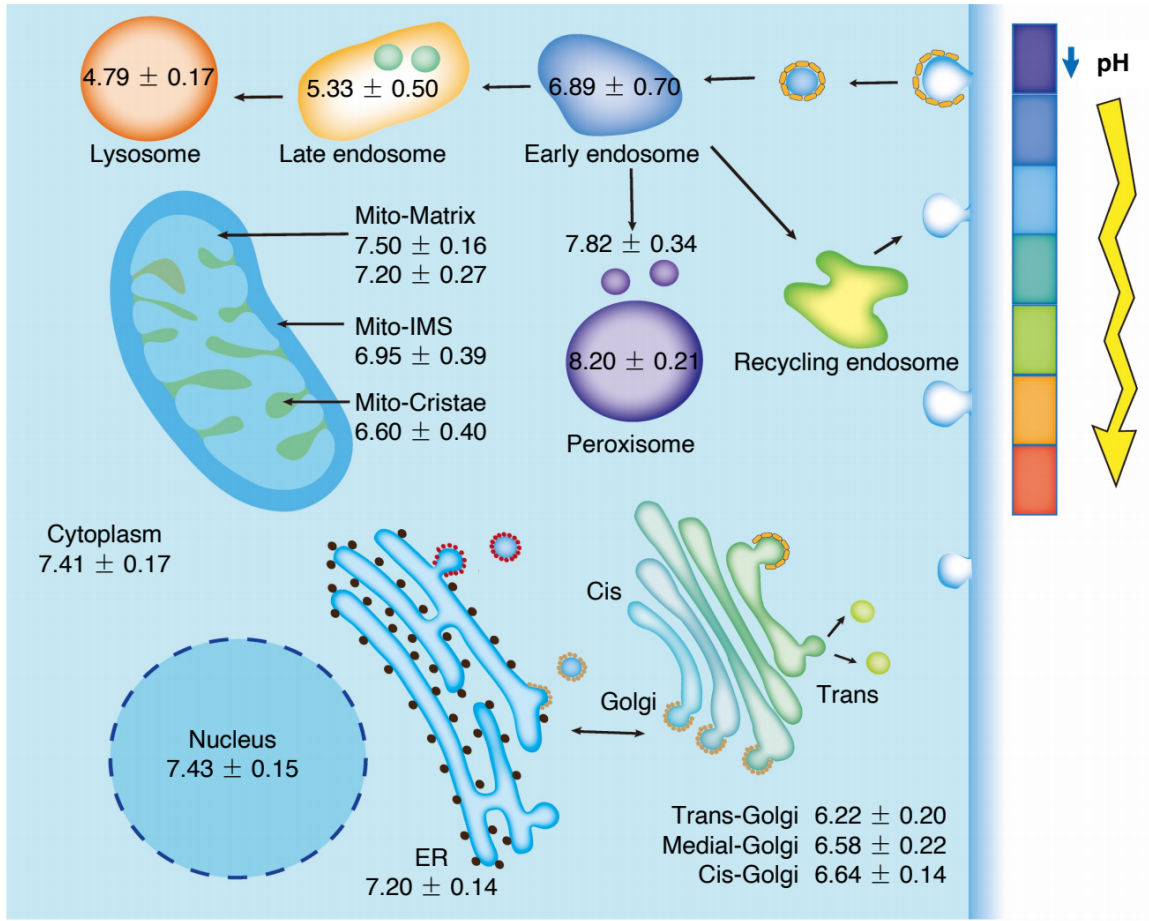

**Fig. 7| The illustration of the pH map of intracellular organelles.**

## Discussion

**Connotations of SITE-pHorin applications**

In this study, we demonstrated that SITE-pHorin was a ratiometric and pH-sensitive sensor. SITE-pHorin exhibited several advantages over other fluorescent proteins, including a wide pH response range from pH 3.5 to pH 9.0, enabling pH measurement in most organelles. The major advantage was its single major excitation and dual emissions characteristics, allowing for ratiometric measurements to quantitatively determine organelle pH. Indeed, we successfully utilized SITE-pHorin conjugated with organelles-targeted markers to measure the pH of various organelles and their sub-compartments, including mitochondrial sub-compartments, Golgi sub-stacks, endoplasmic reticulum, lysosomes, peroxisomes and endosomes.

Significantly, we achieved a small but critical step by successfully determining the pH value and gradients of mitochondrial cristae using the SITE-pHorin probe designed with the g-subunit, based on the topological information in the recently published structure of the ATPase(Gu et al., 2019). Previous findings reported that ΔpH difference across the mitochondrial inner membrane was 0.4-0.6 pH units(Kamo et al., 1979). Our measurement of ΔpH was 0.6-0.9 pH units, which suggested that the prior knowledge about ΔpH was underestimated, likely due to uncorrected noise and autofluorescence. The corrected ΔpH measurement suggested that the proton motive force of mitochondria was also underestimated since ΔpH was an important contributor.

In addition, our investigation intriguingly unveiled that TFR1-SITE-pHorin labeled endosomes distributed into at least three compartments with distinct pH values (Fig. 6b). The pH measurements suggested the presence of diverse TFR1 positive endosomes, such as acidic endosomes (early endosome and late endosome) and lysosomes; on the other hand, based on the fitted results and the pH map (Fig. 7a), we speculated that the unusual alkaline endosomes labelled with TFR1 might potentially interact with peroxisomes and mitochondria. In the near future, SITE-pHorin could enable *in vivo* imaging and quantitative detection of pH variations in tissues and whole organisms under various physiological and pathological states, thereby providing deeper insights into the role of pH regulation in biological processes and uncovering key factors in disease onset and progression. Nevertheless, further studies are warranted to validate and elucidate these speculations.

**The quantum entanglement mechanism in quantum chemistry of SITE-pHorin**

The most unexpectable phenomenon was quantum entanglement, which emerged from the structural evolution and optimization of fluorescent proteins. The Fig. 4 demonstrated that the protonation/deprotonation of the phenolic hydroxyl moiety in Y182 was tightly and oppositely coupled with the status of the phenolic hydroxyl moiety in the chromophore of SITE-pHorin, and vice versa. Consequently, such a tight coupling mechanism via quantum entanglement, enabled the chromophore of SITE-pHorin to sense the pH change in its surrounding environment via the pH-sensitive residue (the outside Y182 residue), resulting in the pH ultra-sensitivity of SITE-pHorin.

The quantum entanglement mechanism of the pH ultra-sensitivity of SITE-pHorin remains unclear. We conducted advanced computational analyses using density-functional theory (DFT) with NWChem software (Valiev et al., 2010) within the B3LYP/6-31G** computational framework, a widely recognized method in quantum chemistry. This approach allowed us to calculate the interaction energies between the Y182 residue and the chromophore under distinct pH conditions. Specifically, we explored low and high pH environments to comprehensively evaluate the impact of protonation/deprotonation changes on the interaction energies (Table S3). As a control, we examined GFP-S65T, which also displayed non-trivial deprotonation of its Y182 residue (Fig. S10). To our surprise, under low pH, the deprotonated Y182 residues of SITE-pHorin and GFP-S65T were sufficient to induce interaction, while protonated Y182 residues under high pH showed little interaction with the chromophores (Table S4). Clearly, the deprotonation of Y182 alone was not sufficient to trigger quantum entanglement.

To ensure greater accuracy, we recalculated the interaction energy of the chromophore in SITE-pHorin under low pH using a larger basis set (Aug-cc-pVDZ), obtaining a value of -0.3653 Hartree (~9.94 eV). This was close to the energy of benzene ring with a distorted geometry of benzene ring(Casanova and Alemany, 2010), so that we measured the internal angles of benzene rings in Y182s and chromophores of SITE-pHorin and GFP-S65T (Fig. S13a and b). We found that the benzene ring in the chromophore of SITE-pHorin at pH 5.5 exhibited an angle of 719.7°, which deviated by -0.3° from the normal 720.0° of a flat and symmetric benzene ring. Considering the large root-mean-square errors (~0.966-1.938°) of bond angles (Table S2) and the rigidity of benzene structure, we measured, plotted, and fitted the distribution of the internal-angle sums of benzene rings in phenylalanine and tyrosine residues of SITE-pHorin under various pH conditions. Consequently, the sum of these internal angles of benzene rings followed a Gaussian distribution with a fitted peak at 719.98° and a small standard deviation of 0.076°, consistent with the rigidity of benzene ring. Thus, the distribution of the sum of internal-angles of the benzene structure suggested that the 719.7° of the benzene ring in the chromophore of SITE-pHorin at pH 5.5 was significantly (p =

0.000115) deviated from the normal 720.0°, indicating a discrete curvature(Gu et al., 2018) of -0.3°. The discrepancy (-0.0161 Hartree/~0.438 eV) of the interaction energy between SITE-pHorin and GFP-S65T in Table S3 under low pH conditions might be related to the curvature, which is necessary for the quantum entanglement since the deprotonation of Y182 alone was insufficient. In the realm of physics, gravity and Yang-Mills field strength intertwine with curvature forms (Baez and Muniain, 1994). Quantum gravity, once considered feeble, may now be intricately linked to the quantum entanglement mechanism of SITE-pHorin. This potential connection arises from the application of gravity in the context of gauge theory squared (Borsten, 2020). Little is known about quantum entanglement under high pH conditions, though it may involve mechanisms similar to those in low pH environments. Quantum entanglement in SITE-pHorin might likely be initiated during structural maturation after its expression. Therefore, SITE-pHorin, which exhibits pH-regulated quantum properties, is a promising probe not only for biologists but also for chemists and physicists exploring quantum chemistry, quantum mechanics, quantum gravity, and quantum information encoding and communication.

## Methods

**Molecular cloning and mutagenesis**

The pmTurquoise2-Mito plasmid (Addgene, # 36208) generously provided by Dorus Gadella served as the initial template. Using PCR with KOD-Plus polymerase (TOYOBO, KOD-201), mTurquoise2 was cloned into pET28a vector for proteins expression. Site-directed mutagenesis was performed using the QuikChange® II Site-Directed Mutagenesis Kits to introduce mutations at the C48, S65, W66, T203, and D148 sites, yielding single, double, triple and quintuple mutants. These mutants were then cloned into the pET28a vector for proteins expression. The deGFP4 was also generated by introducing mutations (S65T/C48S/H148C/T203C) into the EGFP and cloned into the pET28a vector.

    Determination of organelle-targeted SITE-pHorin constructs involved several steps. SITE-pHorin constructs were initially cloned into pEGFP-N1 vector using the Age I and BsrG I restriction enzyme site. The cDNA for human H2B, human 4Cox8, and rat TFR1 were obtained from the previous Jiansheng-Kang laboratory. Human ATP synthase g subunit, human ST6GAL1, MGAT2, and GP73 cDNA were synthesized in Genewiz company. Mouse Ferrochelatase (Fech), mouse Dnase2B, mouse Calreticulin cDNA were sourced from mouse brain tissue. The SITE-pHorin-SKL construct was created by replacing the last amino acid YK with SKL. The hH2B, hComplex V g subunit, four-repeat Cytochrome c oxidase subunit 8 (4Cox8), mFech, hST6GAL1, hMGAT2 (1-89), hGP73 (1-34), mCalreticulin, mDnase2B, and SITE-pHorin-SKL were cloned into the SITE-pHorin-N1 vector. Details of the organelle-targeted genes were summarized in Table S2. All plasmids were sequenced by GeneWiz before further analysis. Q5 polymerase, restriction enzymes and T4 DNA ligase were purchased from New England Biolabs.

**Protein expression, purification, and crystallization**

For spectra measurements, the mTurquoise2 wild-type and all the mutant proteins were expressed in *Escherichia coli* strain BL21(DE3) using the pET28a expression vector with N-terminal 6×His tag. Bacteria were cultured with 0.3 mM IPTG and 18°C overnight. After centrifugation, bacterial pellets were resuspended in 20 mM Tris-HCl, pH 7.5, 300 mM NaCl, 10 mM β-ME, 0.1 mM PMSF. Cell lysates were obtained by sonication for 10min (200w, work time 1s, interval 2s). The lysates were then centrifuged, and Ni-agarose resin was added to the supernatant and rotated at 4°C for 1h. The Ni-agarose-bound proteins were washed three times with 20 mM Tris-HCl pH 7.5, 300 mM NaCl, 20 mM Imidazole, 10mM β-Mercaptoethanol (β-ME). Elution of the proteins was performed with three column volumes of 20 mM Tris-HCl pH 7.5, 300 mM NaCl, 200 mM Imidazole, 10mM β-ME. The eluted proteins were concentrated to 2 mL using an ultrafiltration tube (Millipore, UFC201024). The buffer was exchanged to PBS by PD10 columns (GE, 17-0851-01). Protein concentration was determined using a BCA kit and adjusted to 5 mg/mL.

For crystallization, mTurquoise2 S65T, W66Y, SITE-pHorin and SITE-pHorin_C203 E were purified using affinity chromatography with HisTrap HP (5 ml) column, followed by gel-filtration with HiLoad 10/300 Superdex column. The affinity purification involved the use of the following buffers: Buffer A: 50 mM Tris-HCl pH 8.0, 300 mM NaCl; Buffer B: 50 mM Tris-HCl pH 8.0, 300 mM NaCl, 20 mM Imidazole; Buffer C: 50 mM Tris-HCl pH 8.0, 300 mM NaCl, 250 mM Imidazole. The gel-filtration buffer contained 10 mM Tris-HCl pH 8.0, 100 mM NaCl, 1mM DTT. mTurquoise2 S65T was concentrated into 13 mg/mL and crystallized under conditions of 20% PEG 8000, 100 mM $MgCl_2$, 100 mM HEPES pH 6.5. mTurquoise2 W66Y was concentrated to 15 mg/mL and crystallized under conditions of 10% PEG 8000, 100 mM MgCl$_2$, 100 mM HEPES pH 6.5. SITE-pHorin was concentrated to 10 mg/ml and crystallized under conditions of 14% PEG 8000, 100 mM MgCl$_2$, 100 mM HEPES pH 7.0 and 0.1 M Bis-tris pH 5.5, 25% PEG 3350. SITE-pHorin_C203E was also concentrated to 10 mg/mL and crystallized under conditions of 20% PEG 6000, 100 mM MgCl$_2$, 100 mM Citrate Acid pH 5.0 and 20% PEG 8000, 100 mM MgCl$_2$, 100 mM Tris-HCl pH 8.0. All crystals were grown at 289 K using the hanging-drop vapor diffusion method.

**Data collection and structure determination**

Diffraction data were collected on beamline BL19U1 at the Shanghai Synchrotron Radiation Facility (SSRF). Collected data were processed by HKL3000. The crystal structures of mTurquoise2 S65T, W66Y, SITE-pHorin (at pH 7.0 and pH 5.5), and SITE-pHorin_C203E (at pH 5.0 and pH8.0) were determined by molecular replacement using the predicted structures from mTurquoise2 as search models. Structure refinement and model building were performed with Refmac, PHENIX and Coot. All models were validated with MolProbity. Details of the data processing and refinement statistics were summarized in Table S1. All structure Fig.s were prepared with PyMOL (https://www. pymol.org).

**Spectra measurements**

The purified mTurquoise2, mTurquoise2 mutants, SITE-pHorin and deGFP4 were used to test the absorption and emission spectra by using a UV-visible spectrophotometer (Agilent,

Cary100). Excitation and emission spectra were measured using a 96-well microplate fluorometer (Thermo, Varioskan Flash). For pH titration experiments, solutions were prepared in a series of buffers ranging from pH 3.5 to pH 9.0, consisting of 125 mM Potassium gluconate, 20 mM Sodium gluconate, 0.5 mM $CaCl_2$, 0.5 mM $MgCl_2$, 25 mM of one of the following components: citric acid (pH 3.5, 4.0), acetic acid (pH 4.5, 5.0), MES (pH 5.5, 6.0), PIPES (pH 6.5), HEPES (pH 7.0), Tricine (pH 7.5), Tris (pH 8.0, 8.5), CHES (pH 9.0). The quantum yield of fluorescence ($\Phi$) of the proteins was calculated relative to deGFP4 which had equal optical density and known values ($\Phi L$ =0.08, $\Phi H$ =0.15).

**Cell transfection, and imaging**

The Cos-7 cells were plated on 12 mm coverslips coated with Matrigel (Coring, 356234) and placed in 35 mm dishes at a density of $3.0$-$5.0 \times 10^5$ cells per dish. After 24 hours of plating, the SITE-pHorin-N1 and organelle-targeted SITE-pHorin plasmids were transfected into COS7 cells using the calcium phosphate transfection method following the manufacturers' instructions.

The standard curve of pH-sensitive SITE-pHorin was established both in vitro and in vivo. For the in vitro measurements, the purified SITE-pHorin protein was concentrated to 1 mg/mL and imaged using a confocal microscope (ZEISS, LSM 980) with a single excitation wavelength of 405 nm and two emissions wavelength channels: 420-480 nm and 500-570 nm. The ratio of fluorescent intensities (log (Fb/Fa)) was calculated at pH values ranging from 3.5 to 9.0 to create the standard curve. The standard curve for in vivo measurements was obtained using Cos7 cells that were transfected with SITE-pHorin-N1 plasmids. The transfected Cos7 cells were initially treated with Nigericin/Valinomycin at the same concentration as the Intracellular pH Calibration Buffer Kit (ThermoFisher, P35379). Subsequently, imaged using the confocal microscope was employed to image the cells using with the same excitation and emission wavelength mentioned above, in buffers ranging from pH 3.5 to 9.0. The standard curve in vivo was also be created by the ratio of fluorescent intensities (log (Fb/Fa)) of Cos7 cells at pH values ranging from 3.5 to 9.0. The pH of the organelle-targeted SITE-pHorin was then calculated using the standard curve in vivo (Fb, the fluorescence intensity of the emission wavelength 500-570 nm was measured with the excitation wavelength 405 nm; Fa, the fluorescence intensity of the emission wavelength 420-480 nm was monitored with the excitation wavelength 405 nm).

**The quantum mechanism of the SITE-pHorin**

For a single base-acid equilibrium,

$$H^+ + Base(B) \overset{K}{\leftrightarrow} Acid(A) \tag{I}$$

where *K* is the value of the acid dissociation constant; *A* represents the protonated form; *B* is the conjugate base of *A*.

According to the Henderson-Hasselbalch equation (I), we have:

$$pH = pK_a + log_{10}([B]/[A]) \tag{II}$$

where *pK$_a$* is the value of the acid dissociation constant; *[B]* and *[A]* represent the concentrations of base and acid forms respectively.

The slope of the relationship between the logarithm of the fluorescent signal ratio and the pH is one for a single acid in the equation (II). For a pH sensitive fluorescent protein, let's assume that the protein needs n simultaneous protonations in or nearby its chromophore for the pH sensitivity, and that the system is at equilibrium, the ratio of deprotonated to protonated forms can be expressed as follow:

$$\frac{[B]}{[A]} = \frac{K}{[H^+]^n} \qquad (II)$$

Then, we obtain the ratio of the fluorescent intensities between base and acid forms for this multi-protons hypothesis:

$$\frac{F_b}{F_a} = \frac{[B] \cdot \Phi_b}{[A] \cdot \Phi_a} \cdot g = \frac{K}{[H^+]^n} \cdot \frac{\Phi_b}{\Phi_a} \cdot g \qquad (IV)$$

where $F_a$ and $\Phi_a$: the fluorescence and the fluorescent quantum yield of the acid form; $F_b$ and $\Phi_b$: the fluorescence and the fluorescent quantum yield of the base form; $g$ is a constant for an imaging acquiring setup.

Taking the negative logarithm of both sides of the equation (IV) and rearranging, the multi-protons hypothesis gives the following expression:

$$pH = \frac{1}{n} \cdot log_{10}\left(\frac{F_b}{F_a}\right) + \frac{1}{n} \cdot (pK_a - log_{10}(\frac{\Phi_b}{\Phi_a} \cdot g)) \qquad (V)$$

If the ratio of fluorescent intensity is normalized and set to one ($log_{10}\left(\frac{\Phi_b}{\Phi_a} \cdot g\right) = (1-n)pKa$) when pH is equal to pKa, the equation (V) is reduced to:

$$pH = \frac{1}{n} \cdot log_{10}\left(\frac{F_b}{F_a}\right) + pK_a \qquad (VI)$$

Clearly, the slope 1/n of the equation (VI) reflects the sensitivity of the protein to the change in pH, which is just the assumed fact that n simultaneous protonations need for its pH sensitivity.

Interestingly, the pH sensitive curve normalized with the value around the pKa (≅6) of the SITE-pHorin shows that a fitted slope for a physiological pH range (4.0-8.0) is around 2 (Fig. S11), which suggests the pH sensitivity of the SITE-pHorin is double the sensitivity of a single protonation model. The intercept is just the pKa, so that we fit and get a pH sensitive expression of the SITE-pHorin:

$$pH = 2 \bullet log_{10}\left(\frac{F_b}{F_a}\right) + pK_a \tag{VII}$$

Apparently, the multi-protons model couldn't explain this large slope. In addition, there are no multiple protonations in or near the chromophore of the SITE-pHorin (Fig. 4).

Out of expectation, the real situation is astonishing since there are simultaneously opposite protonation and deprotonation between the chromophore of the SITE-pHorin and the residue Y182, which is located outside of the β-barrel structure (Fig. 4). The results suggest the protonation/deprotonation of the phenolic hydroxyl moiety in the Y182 residue is tightly and opposite coupled with the protonation/deprotonation of the phenolic hydroxyl moiety in the chromophore of the SITE-pHorin, and vice versa. Therefore, such coupling mechanism enables the chromophore of the SITE-pHorin coupling the environment outside of the barrel via the Y182 residue to sense the change of pH.

Straightforwardly, the slope might be explained as a quantum effect (n = 1/2) for its pH ultra-sensitivity of the SITE-pHorin, which could be enhanced by the quantum entanglement (the opposite and simultaneous protonation and deprotonation) between the phenolic hydroxyl moieties of the chromophore and Y182 (Fig. 4). The quantum effect and quantum entanglement in the pH ultra-sensitivity of the SITE-pHorin might just reflect the characterization of holonomy and non-trivial topology of electron rotations(Puentes, 2017) (*61*) in the resonance structures of benzene rings embedded in the two phenolic hydroxyl moieties (Fig. 4).

Consequently, the quantum entanglement explanation could properly fit the structural phenomenon (Fig. 4) and the fluorescent data in the physiological pH (4.0-8.0) (Fig. S12), and well explain the mechanism and pH ultra-sensitivity of the SITE-pHorin. On the other hand, there were some less well fitted data points, such as a high pH at 9.0 or a very low pH at 3.5. The low pH at 3.5 might partially be due to the stability of protein at such low pH although the fluorescent signal of the SITE-pHorin was stable and pretty good in lysosome (Fig. 5). The high pH at 9.0 might interfere the quantum entanglement between the phenolic hydroxyl moieties of the chromophore and Y182 since the quantum entanglement seemed to be stronger (a shorter distance between the phenolic hydroxyl moieties) in a lower pH (Fig. 4), which might also partially contribute to the stability of the SITE-pHorin in acid environments.

**Background noise and autofluorescence corrections for pH measurements**

Fluorescent images usually contain background noise or autofluorescent signals, which inevitably affect the accuracy of the measurements based on fluorescent intensity. Thus, it is necessary to know how noise or autofluorescence will affect the results and how to correct those potential distortions.

Fluorescent intensity ($F$) contains a real signal ($r$) and a noise/autofluorescence interference ($n$), and thus the signal-to-noise ratio ($snr$) is $r/n$.

$$F = r + n, and\ snr = r/n \tag{VIII}$$

Then substituting this equation (VIII) into the logarithm of the fluorescent signal ratio, we obtained:

$$log_{10}\left(\frac{F_b}{F_a}\right) = log_{10}\left(\frac{r_b+n_b}{r_a+n_a}\right) = log_{10}\left(\frac{r_b}{r_b}\right) + log_{10}\left(\frac{1+n_b/r_b}{1+n_a/r_a}\right)$$

Rearranging the equation, we get:

$$log_{10}\left(\frac{r_b}{r_a}\right) = log_{10}\left(\frac{F_b}{F_a}\right) - log_{10}\left(\frac{1+1/snr_b}{1+1/snr_a}\right) \tag{IX}$$

Clearly, to acquire a real the logarithm of the fluorescent signal ratio, we just need calculate and adjust the measurement with the second term in the equation (IX), which is all about the *snr* of imaging channels. The second logarithm term of the equation (IX) is linearly separated from the logarithm of the real signal ratio, which makes the correction easy and efficient. Moreover, the equation (IX) also suggests that if imaging channels have comparable *snr*, the pH measurement will be free of the influence by background noise/autofluorescence, which is the favorable advantage and reliability of ratiometric measurement. In our experiments, the adjustments of the pH measurements in the most of cases were in the range of 0.003-0.07.

If both imaging channels have a very good *snr*, the Taylor expansion of the second term in the equation (IX) approximates to

$$log_{10}\left(\frac{1+1/snr_b}{1+1/snr_a}\right) \cong log_{10}(e)\left(\frac{1}{snr_b} - \frac{1}{snr_a}\right) \qquad (X)$$

Using the equations (VII), (IX) and (X), we obtain:

$$pH \cong pK_a + 2log_{10}\left(\frac{r_b}{r_a}\right) + 2log_{10}(e)(1/snr_b - 1/snr_a) \qquad (XI)$$

Consequently, the equation (XI) can tell us that a noisy *snr* of the base channel (515 nm peak) leads to overestimate the pH value, and while a noisy *snr* of the acid channel (465 nm peak) results in underestimating the pH measurement. A worse case could have a superimposed and distortional effect from multiple noisy channels.

The equation (XI) demonstrates how noise or autofluorescence affects the results. The situation could be severe where the signals were mixed with strong autofluorescence, such as mitochondrial sub-compartments. What's troubling is that the adjustments of autofluorescence usually are not possible done in the same experiments as the corrections for background noise.

Thanks to the logarithm, the second logarithm snr term of the equation (IX) for the correction can be fortunately linearly separated from the logarithm of the real signal ratio. To minimize the errors of pH measurements, it is necessary and feasible to do independent control experiments to measure and evaluate the parameters of noise and autofluorescence under the same conditions for an accurate estimation of snr. Especially, the intensities of FAD autofluorescence and NADH autofluorescence are strong so that the autofluorescent ratio of FAD and NADH is frequently used as a measurement of mitochondrial redox potential, so that we used the ratio previously for representing the dynamic of mitochondrial metabolism(Xie et al., 2017). Thus, minimizing the effect of autofluorescence is particularly important for the pH accuracy of the mitochondrial sub-compartments. In our experiments, the corrections of the pHs of mitochondrial sub-compartments were in the range of 0.07-0.44, which were an order of magnitude higher than the adjustments of other organelles due to the autofluorescent interferences of mitochondrial metabolites NADH and FAD.

**Supplemental material**

The supplemental material contains 13 supplemental figures and 4 supplemental tables.

**Data Availability Statement**

The crystal structure has been uploaded to Protein Data Bank database. mTurquoise2 S65T (PDB ID: 8IYZ), mTurquoise2 W66Y (PDB ID: 8IZ0), SITE-pHorin_pH7.0 (PDB ID: 8IYY), SITE-pHorin_pH5.5 (PDB ID: 8IZ3), SITE-pHorin C203E_pH5.0 (PDB ID: 8IZ1), SITE-pHorin C203E_pH8.0 (PDB ID: 8IZ2). All other data are available in the article itself and its supplementary materials.

## Acknowledgements

We thank the facilities in the Clinical Systems Biology Laboratories and the Translational Medicine Center of the First Affiliated Hospital of Zhengzhou University, and the facilities in the Shanghai Institutes for Biological Sciences, Chinese Academy of Sciences. Thank Dorus Gadella for sharing the pmTurquoise2-Mito plasmid.

**Funding:** National Natural Science Foundation (NSF) of China: JSK (92054103, 32071137); Funding for Scientific Research and Innovation Team of The First Affiliated Hospital of Zhengzhou University: JSK (ZYCXTD2023014); National Natural Science Foundation (NSF) of China: SAL (32000855); National Natural Science Foundation (NSF) of China: PPL (32000522); Joint Construction Program for Medical Science and Technology Development of Henan Province of China: SAL (LHGJ20190239); Joint Construction Program for Medical Science and Technology Development of Henan Province of China: PPL (2018020088); Natural Science Foundation of Henan Province of China: PPL (202300410420).

## Author contributions

Conceptualization, J.S.K.; Methodology, J.S.K.; Software, J.S.K; Investigation, S.A.L, X.Y.M., S.Z., R.Z.Y., Y.J.Z., D.D.W., P.P.L.; Visualization, J.S.K., S.A.L., Y.Y.; Funding acquisition, J.S.K., S.A.L., P.P.L.; Project administration, J.S.K.; Supervision, J.S.K.; Writing – original draft, S.A.L., J.S.K.; Writing – review & editing, J.S.K.

**Fig. 1| The spectral characteristics of mTurquoise2, mTurquoise2 S65T and W66Y mutants. a** The normalized fluorescence of mTurquoise2 was detected at the emission wavelength 550 nm, and **b** the emission spectrum of mTurquoise2 with an excitation wavelength 450 nm in the buffers with a pH from 3.5 to pH 9.0. **c** The normalized fluorescence of mTurquoise2 S65T mutant at the emission wavelength 550 nm, **d** and the emission spectrum excited at 395 nm in varied pH buffers. **e** The normalized fluorescence of mTurquoise2 W66Y mutant at the emission wavelength 550 nm, and **f** the emission spectrum excited at 365 nm in the same buffer above. **g** The maximal peaks of excitation and emission, the full width at half maximum (FWHM), and the pH for the maximum emission were summarized for mTurquiose2, S65T, and W66Y. **h** The pH sensitive curves were normalized with the peak values of the emission spectrum at 475 nm (mTurquiose2), 475 nm (mTurquiose2 S65T) or 510 nm (mTurquiose2 W66Y).

**Fig. 2| The key residu D148 of mTurquoise2 and its saturation mutations. a** The cartoon structures of mTurquoise2 S65T and W66Y mutants demonstrated that the D148 located in a loop of the mTurquoise2 W66Y but in a β-sheet of the mTurquoise2 S65T. **b** The alignment structure of mTurquoise2 S65T and W66Y displayed that the chromophore (S65-Y66-G67) of mTurquoise2 W66Y formed hydrogen bonds with aspartic acid 148 (D148) and N121, while the chromophore (T65-W66-G67) of mTurquoise2 S65T formed hydrogen bond with S205. **c** and **d** The normalized excitation (dash lines) and emission (solid lines) spectra of the double mutant S65T/W66Y (**c**) of mTurquoise2 and the triple mutant S65T/W66Y/D148C (**d**) were excited with the wavelength of 400 nm or measured at the emission of 550 nm, respectively, in the pH buffers from 3.5 to 9.0. **e** and **f** The normalized pH sensitive curves of the mTurquoise2 with five mutants, which contained C48S/S65T/W66Y/T203C/D148K (**e**) or C48S/S65T/W66Y/ T203C/D148S (**f**), were also measured with the excitation wavelength 400 nm (dash lines) or the emission wavelength 550 nm (solid lines) respectively. **g** and **h** The representative fluorescent intensities of 20 mTurquoise2 mutants C48S/S65T/W66Y/T203C with an extra saturation mutation of D148, were measured with the emission wavelength 550 nm (**g**) or with the excitation wavelength 400 nm (**h**) in the buffer of pH 6.0. The optimum mutant is D148G dubbed SITE-pHorin, the peaks of which are labeled in the excitation spectrum (**g**) and the emission spectrum (**h**), respectively.

**Fig. 3| The pH sensitive characteristics of SITE-pHorin. a** and **b** The fluorescent intensities of the SITE-pHorin (**a**) and deGFP4 (**b**) were measured with the excitation wavelength 400 nm in pH buffers from 3.5 to 9.0, respectively. **c** and **d** The excitation spectra of the SITE-pHorin (**c**) and deGFP4 (**d**) were measured with the emission wavelength 515 nm in the pH-varied buffers. **e** The logarithms of the ratio of fluorescent intensities of the two emission peaks at 515nm and 465nm in the pH-varied buffers were plotted for the comparison between SITE-pHorin (blue) and deGFP4 (red) both excited at 400 nm. **f** The spectral characteristics of SITE-pHorin and deGFP4 were summarized ($\lambda_{ex}$: the major excitation peak; $\varepsilon$: extinction coefficient determined from the absorbance spectrum at pH 7.2; $\lambda_{emL}$: the lower emission peak; $\lambda_{emH}$: the higher emission peak; $\Phi_L$: the quantum yield of $\lambda_{emL}$ measured at pH 5.0; $\Phi_H$: the quantum yield of $\lambda_{emH}$ measured at pH 8.0). **g** The SITE-pHorin protein was expressed and purified using *E.coli* BL21(DE3) strains, and then detected its fluorescence intensities under varied pHs from 3.5 to 9.0. For clarity, the ratio values of two emissions (green < 480 nm and red > 500 nm) of images were represented with pseudo color. Please note that the leftmost scale of pseudo color was different with two other scales.

**Fig. 4| The quantum entanglement mechanism of the pH ultra-sensitivity of SITE-pHorin. a-d** Schematic diagrams of hydrogen-bond network (**a** and **c**) and electrostatic surfaces (**b** and **d**) demonstrated the environment around the chromophore in SITE-pHorin under different pHs. The electrostatic surfaces (**b** and **d**) demonstrated a

quantum entanglement phenomenon, which was the simultaneously opposite protonation and deprotonation between the chromophore of the sitepHorin and the residue Y182, which is located outside of the **β**-barrel structure (green sheets). The blue intensity of surfaces represented the strength of positive potential, and the red intensity represented the distribution of negative potential. The proposed hydrogen bonds networks and the electrostatic potential around the chromophore and the residue Y182 of SITE-pHorin under pH 5.5 (**a**, **b**) or pH 7.0 (**c**, **d**), respectively. **e-g** The chromophore of a single SITE-pHorin mutant (C203E) was lack of the quantum entanglement (the opposite protonation and deprotonation of the phenolic hydroxyl moiety in different pHs as shown in (**a-d**). **h** The single mutation (C203E) of SITE-pHorin lost the high sensitivity to the change of pH.

**Fig. 5| SITE-pHorin targeting to intracellular organelles and their sub-compartments. a** The SITE-pHorin was overexpressed in COS-7 cell lines, and the fluorescent signals of two emission band passes for the wavelength of 500-570 nm and 420-480 nm were imaged with the excitation of a 405 nm laser in varied pH buffers from 3.5 to 9.0. For clarity, the ratio images were also pseudo-colored presented. **b** The schematic diagram of specific organellar proteins fused with SITE-pHorin: H2B, nucleus marker; SITE-pHorin-SKL, peroxisome marker; Dnase2B, lysosome marker; Calreticulin, endoplasmic reticulum marker; TFR1, endosome marker; g subunit, mitochondria cristae marker; 4Cox8, mitochondria matrix marker; Fech, mitochondria intermembrane space marker; ST6GAL1, trans-Golgi marker; MGAT2, media-Golgi marker; GP73, cis-Golgi marker). **c** The specific organelle-targeted SITE-pHorin were overexpressed in Cos7 cell lines and imaged under the same condition as A. Note that the COS-7 cell transfected with the construct of lysosome-targeted SITE-pHorin only showed green dots (the signal from the emission channel of 420-480 nm), and while the COS-7 cell expressed the TFR1-fused SITE-pHorin demonstrated some green dots, red dots (the emission channel of 500-570 nm) and color-merged yellow dots. Scale bars, 3 μm

**Fig. 6| The pH and distributions of intracellular organelles. a** For the pH measurement of each organelle, the data were collected from ~100 COS-7 cells transfected with each organelle targeted SITE-pHorin. The peaks and standard deviations of the pH measurements of were fitted with single or multiple gaussian fittings for various organelles. The representative images of cytoplasm, the nucleus, and the membrane structure organelles, including the endoplasmic reticulum (ER), Golgi sub-stacks were demonstrated with raw merged images (bottom) or pseudo-colored images (upper). **b** The representative images and the peaks and standard deviations of the pH measurements were showed for the peroxisome, lysosome and TFR1 positive endosomes. Note that the TFR1 positive endosomes could be classified into at least three populations with different pHs ranged from 4.0-9.0. **c** The representative images and the peaks and standard deviations of the pH measurements were showed for the mitochondria cristae, matrix, intermembrane space (IMS). Note that the double gaussian fitting of the pH measurements of the mitochondrial matrix demonstrated two populations with comparable pHs with the cytosol. Scale bars, 3 μm.

**Fig. 7| The illustration of the pH map of intracellular organelles.** The illustration was a summarized map of organellar pH measured with a single pH sensitive probe SITE-pHorin, which possessed an ultra pH sensitivity enhanced by a quantum entanglement effect. As illustrated, the nucleus, the endoplasmic reticulum, the Golgi sub-stacks, the peroxisome, the lysosome, the endosomes and the mitochondrial sub-compartments were labelled with the peaks and standard deviations of their pH measurements. Pseudo colors were used to match intracellular pH gradients.

**Supplemental Figures and Tables**

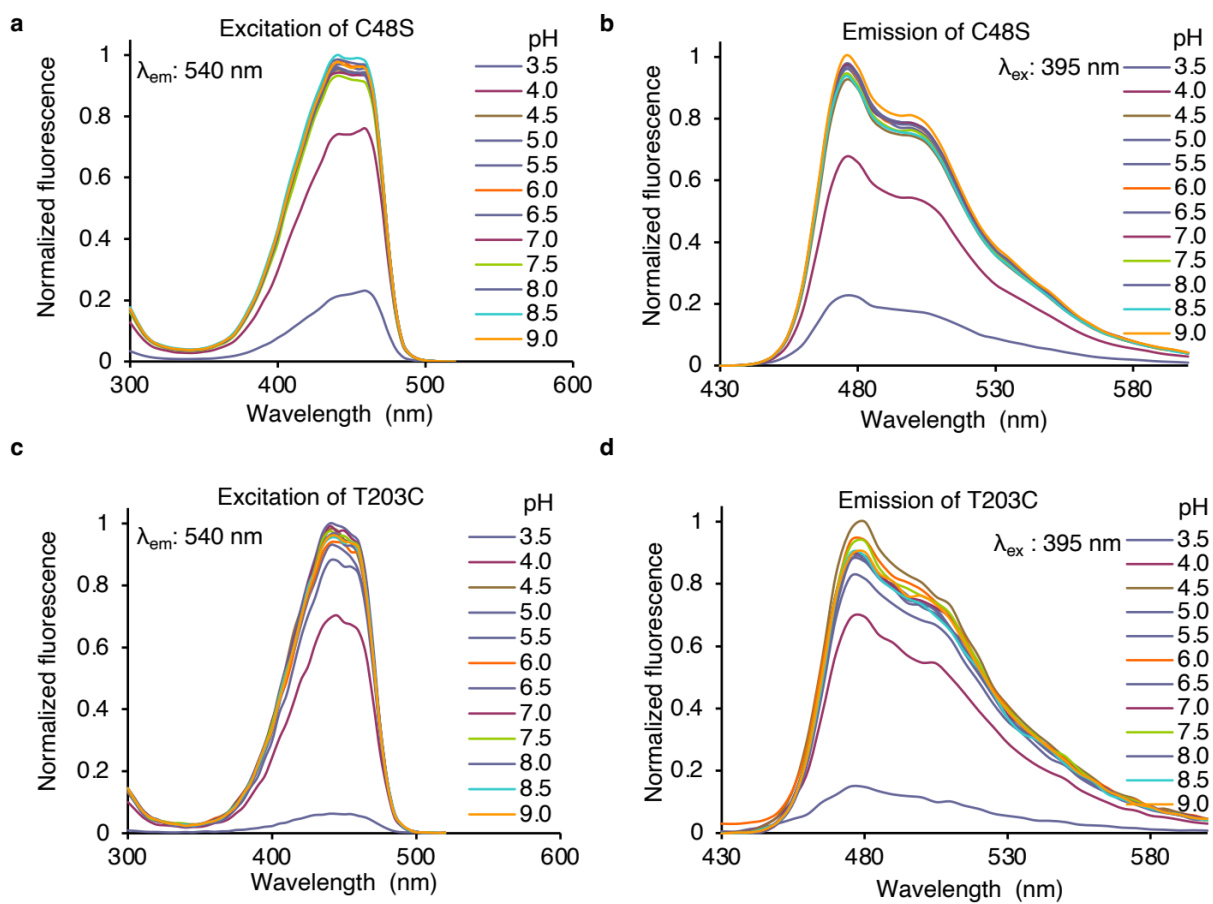

**Supplementary Figure S1. The Spectra of mTurquoise2 mutants with a single mutation C48S or T203C. a** The normalized emission spectrum of the mTurquiose2 mutant C48S was recorded with excitation light at a wavelength 540 nm, and **b** the excitation spectrum of the single mutant C48S which detected at the wavelength 395 nm in varied pH buffers from 3.5 to 9.0. **c** The normalized emission spectrum of the single mutant T203C was analyzed. **d** The normalized excitation spectrum of the single mutant T203C was also determined. The fluorescent data were collected under the same settings as in A or B.

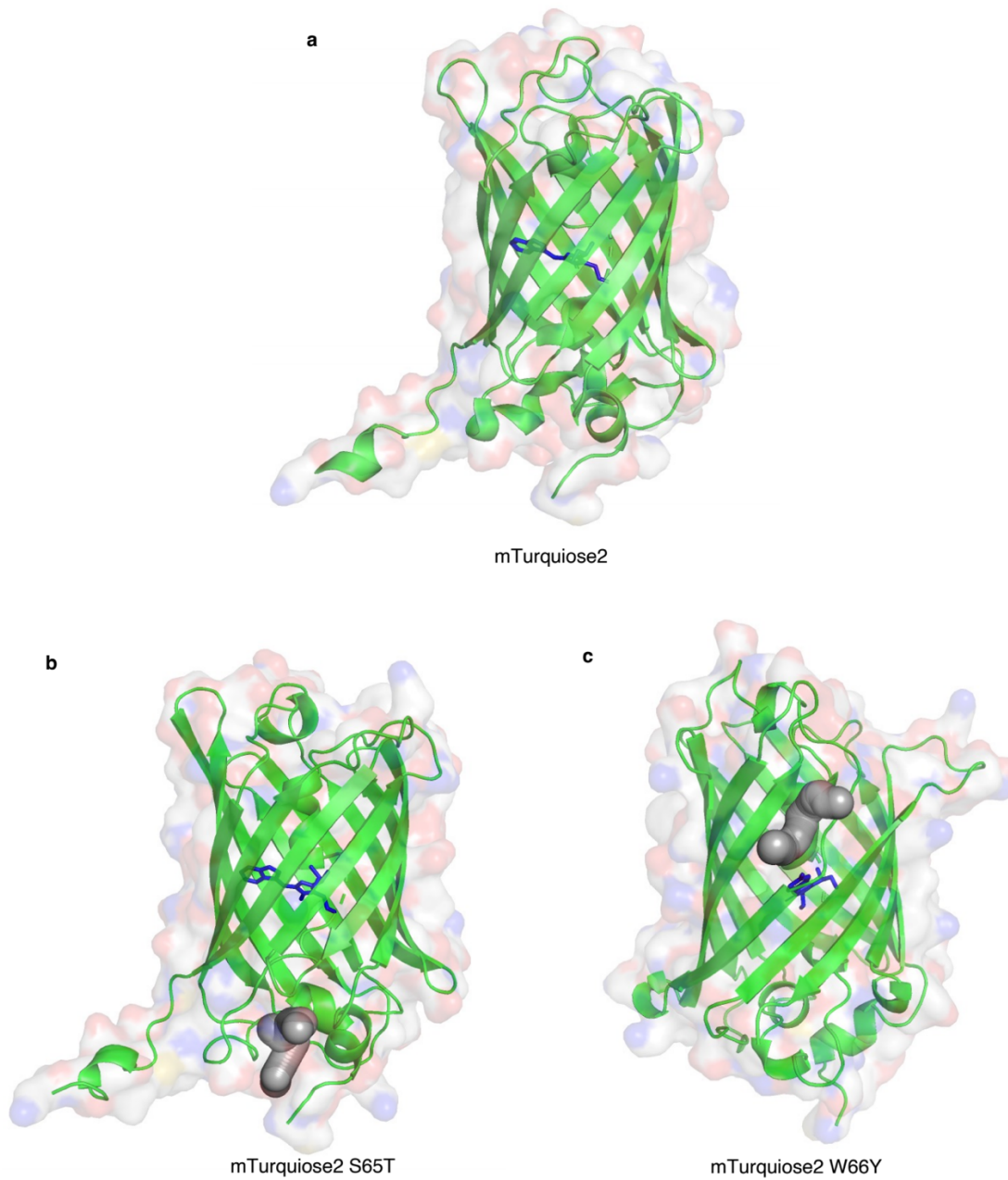

**Supplementary Figure S2. The channels of the mTurquoise2 S65T and W66Y.** The channels of the mTurquoise2 S65T and mTurquoise2 W66Y were analyzed using the MOLE2.5 software. The structures of mTurquoise2 (**a**), mTurquoise2 S65T (**b**) and mTurquoise2 W66Y (**c**) were visualized using PyMOL 2.5.4. The cartoon structures were depicted in green, the surfaces were displayed with a transparency setting of 60%, and the channels of mTurquoise2 S65T and mTurquoise2 W66Y were represented by gray tunnels. The structure of mTurquoise2 (**a**) was lack of tunnels or holes.

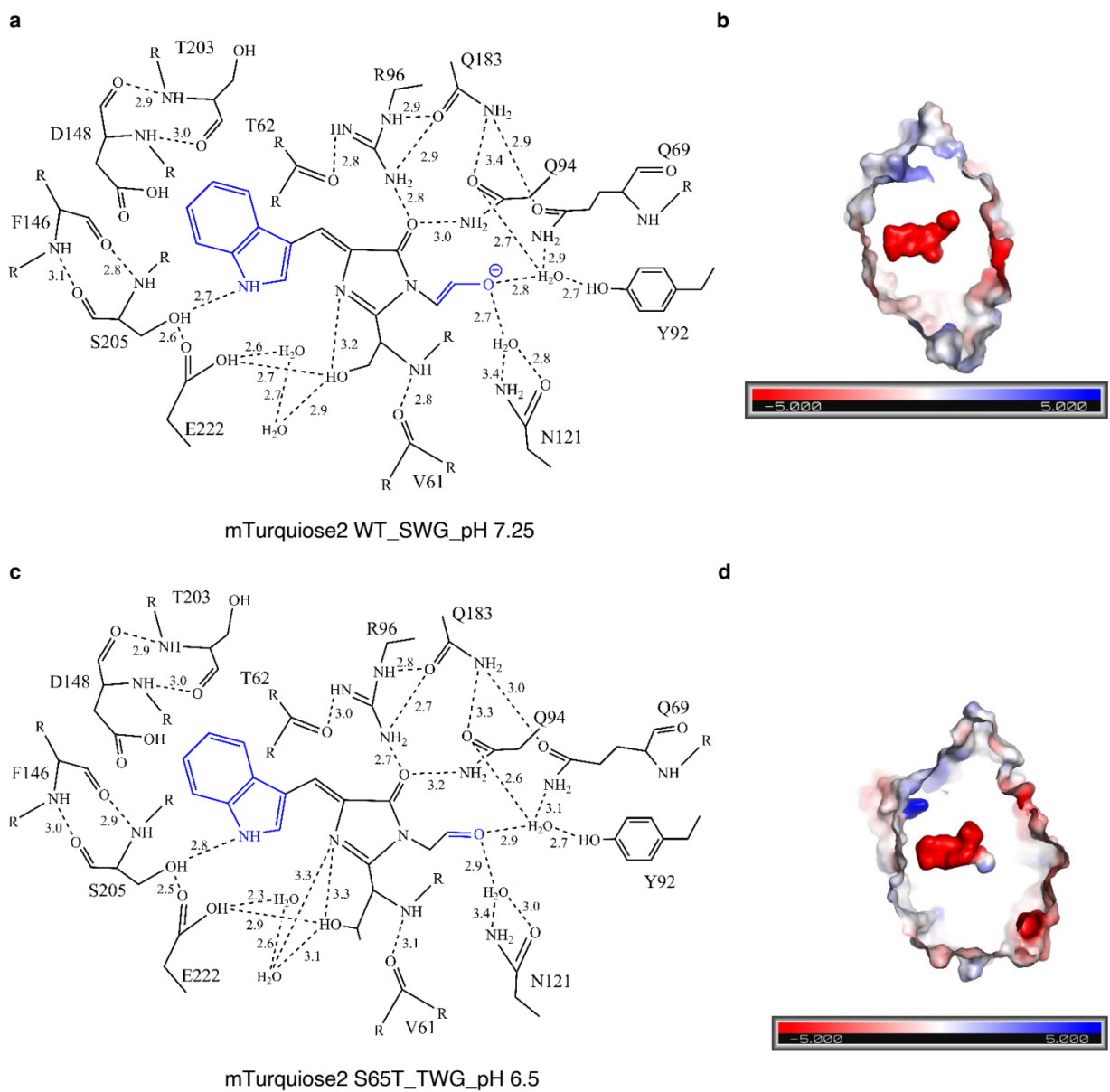

**Supplementary Figure S3. Schematic diagrams illustrating the hydrogen-bond network and electrostatic surfaces around the chromophore of mTurquiose2 and mTurquiose2 S65T. a** Proposed hydrogen bonds of the mTurquiose2 chromophore are depicted as dashed lines. **b** The electrostatic surface of the mTurquiose2 near its chromophore is presented. **c** and **d** Demonstrations of the proposed hydrogen-bonds network (**c**) and the electrostatic surface (**d**) of the mTurquiose2 S65T around its chromophore. The dashed lines represented hydrogen bonds.

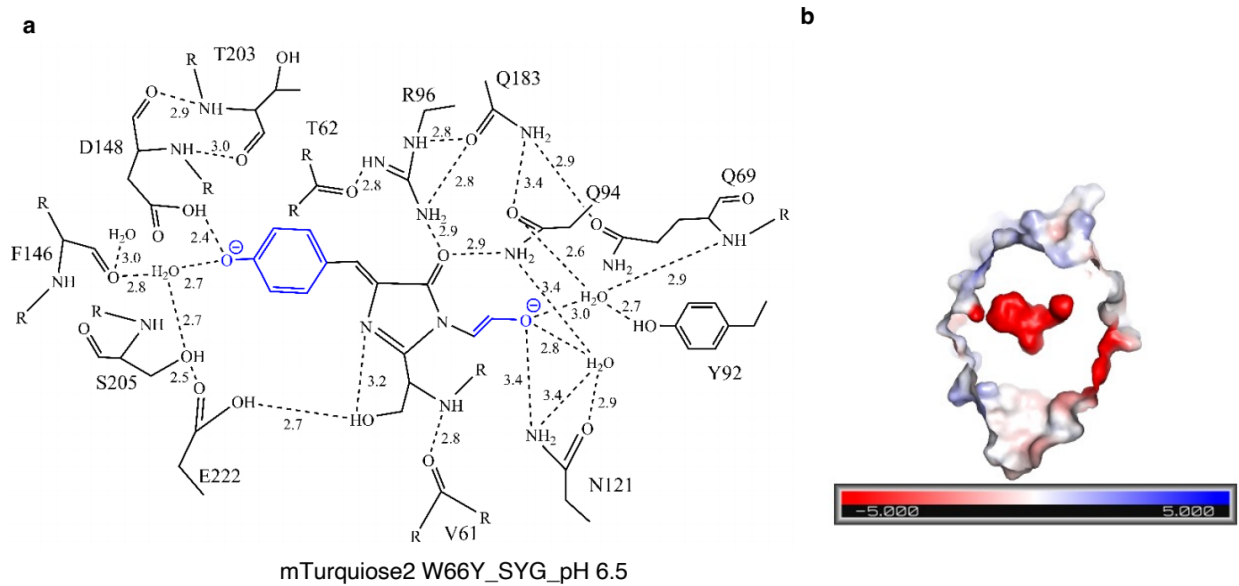

mTurquiose2 W66Y_SYG_pH 6.5

**Supplementary Figure S4. Schematic diagrams illustrating the hydrogen-bond network and electrostatic surfaces around the chromophore of mTurquiose2 W66Y. a** The hydrogen-bonds network and **b** the electrostatic surface of the mTurquiose2 W66Y around its chromophore are shown. The dashed lines represented hydrogen bonds.

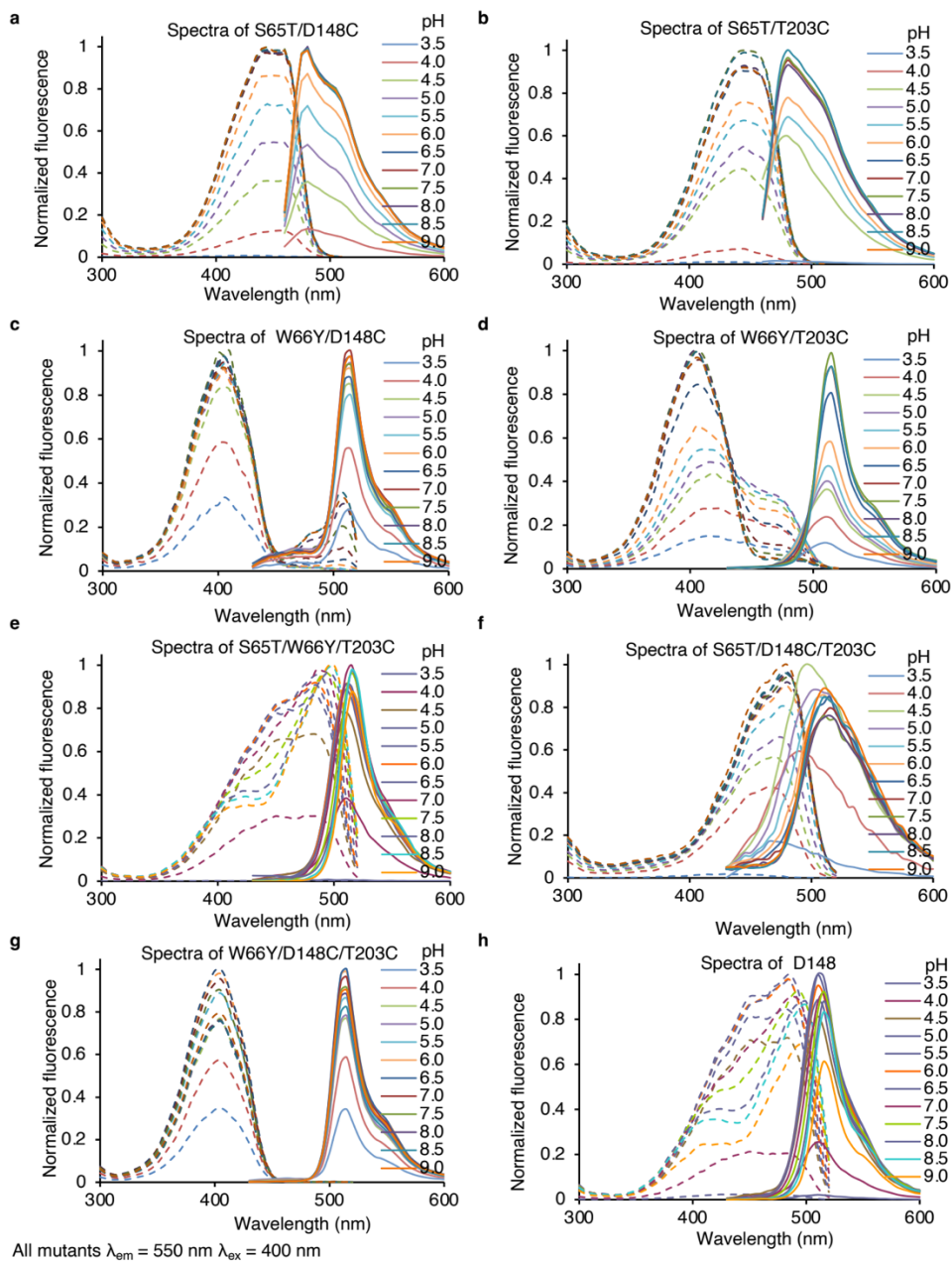

**Supplementary Figure S5. The Spectra of double, triple and quadruple mutants of mTurquiose2. a-d** Normalized spectra of mTurquoise2 double mutants, including S65T/D148C (**a**), S65T/T203C (**b**), W66Y/D148C (**c**), W66Y/T203C (**d**) were plotted. Dashed lines represent excitation spectra detected with the emission wavelength 550 nm, and the solid lines represent emission spectra excited with the wavelength 400 nm in varied pH buffers from 3.5 to 9.0. **e-g** Normalized excitation (dashed lines) and emission spectra (solid lines) of mTurquiose2 triple mutants, including S65T/W66Y/T203C (**e**), S65T/D148C/T203C (**f**) and W66Y/D148C/T203C (**g**) were demonstrated. **h** The normalized spectra of the mTurquiose2 quadruple mutant containing C48S/S65T/W66Y/T203C.

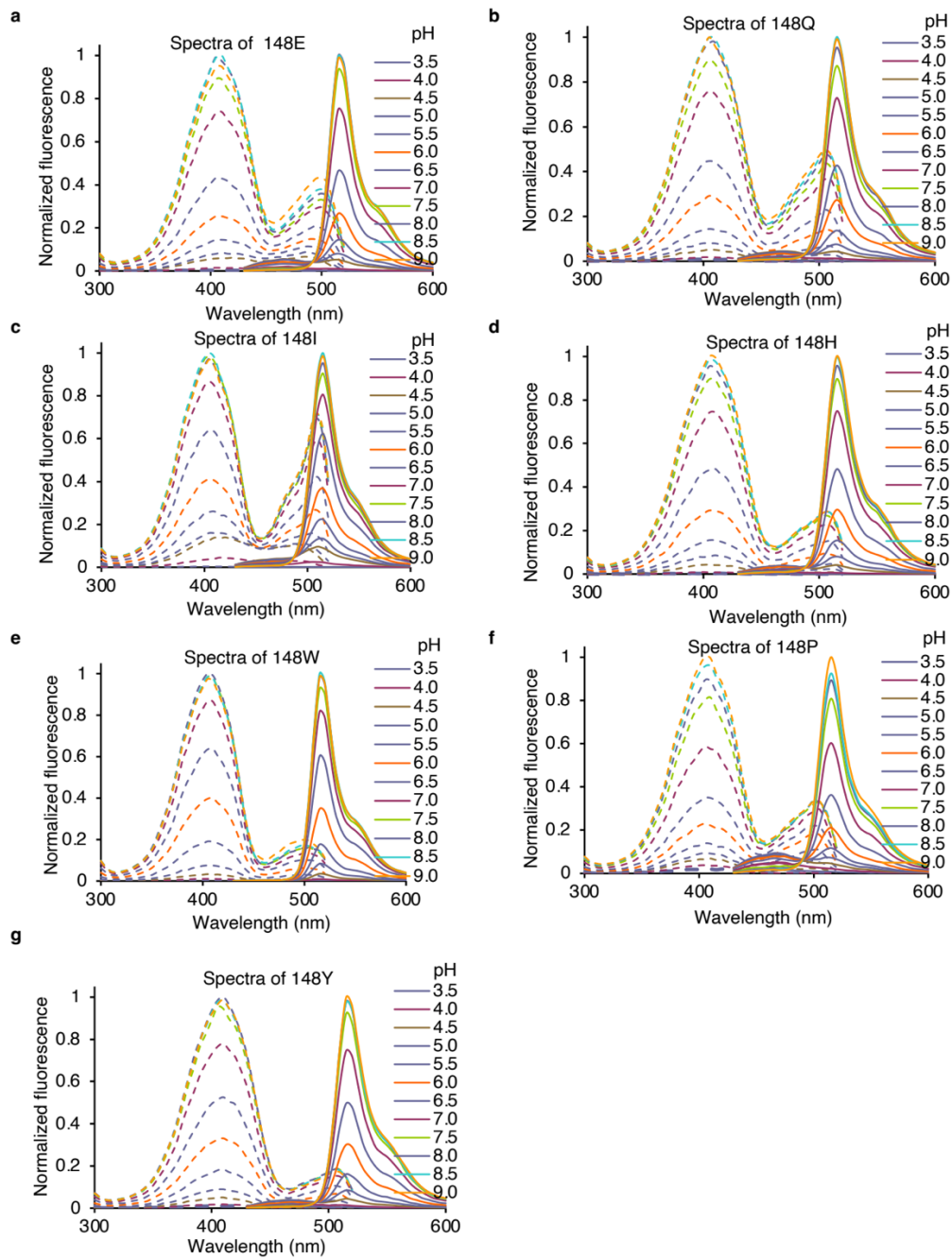

**Supplementary Figure S6. SITE-pHorin G148 mutants with two excitations (the major peak at 400 nm) and two emissions. a-g** SITE-pHorin 148 mutants, including D148E (**a**), D148Q (**b**), D148I (**c**), D148H (**d**), D148W (**e**), D148P (**f**) and D148Y (**g**) were depicted. The excitation spectra (dashed lines) were detected with the emission wavelength 550 nm and the emission spectra (solid lines) was measured with the excitation wavelength 400 nm in varied pH buffers ranging from 3.5 to 9.0.

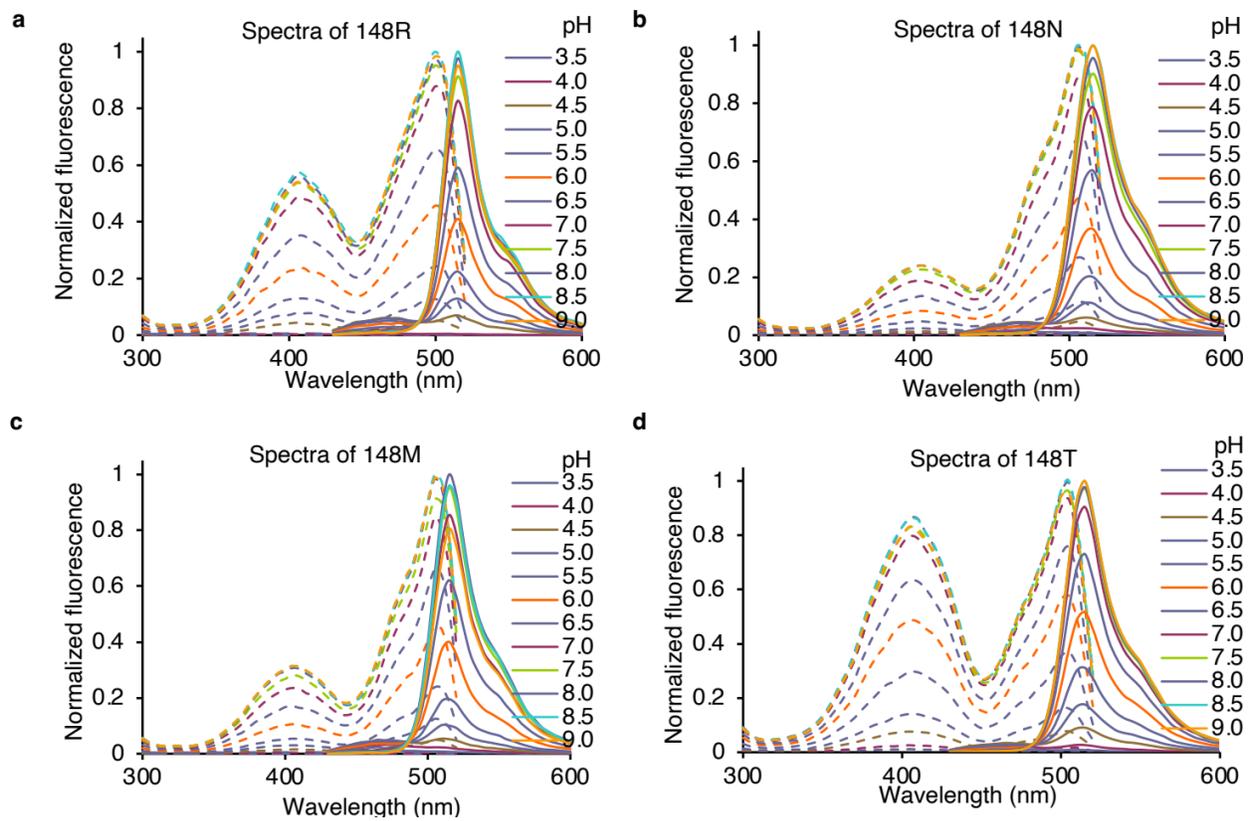

**Supplementary Figure S7. SITE-pHorin G148 mutants with two excitations (the major peak around 500 nm) and two emissions. a-d** SITE-pHorin 148 mutants, including 148R (**a**), 148N (**b**), 148M (**c**) and 148T (**d**) were exhibited. The excitation spectra (dashed lines) were detected with the emission wavelength 550 nm and the emission spectra (solid lines) was measured with the excitation wavelength 400 nm in varied pH buffers ranging from 3.5 to 9.0.

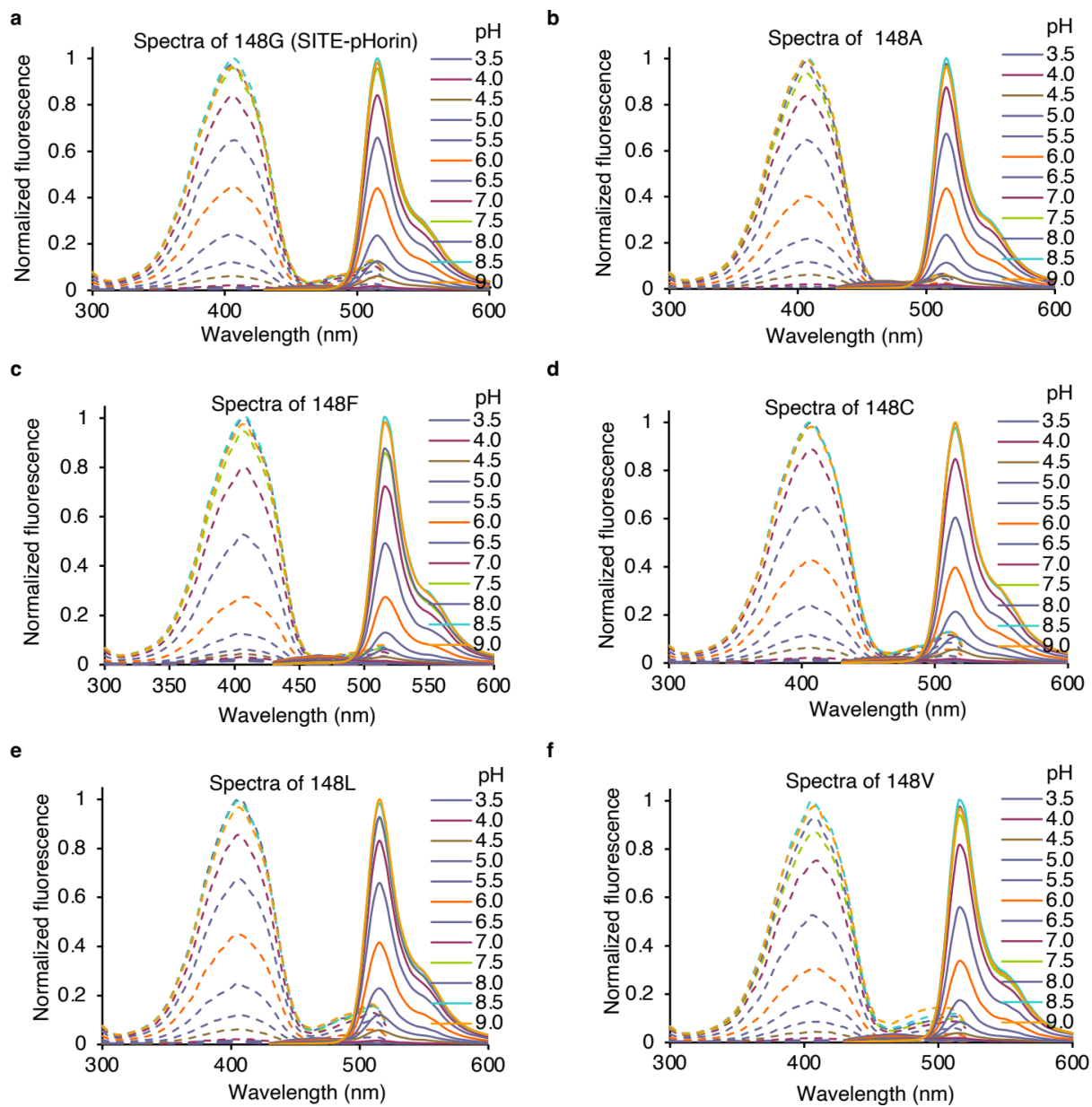

**Supplementary Figure S8. SITE-pHorin and its 148 mutants with one major excitation and two emissions. a-f** SITE-pHorin 148 mutants, including 148G (SITE-pHorin) (**a**), 148A (**b**), 148F (**c**), 148C (**d**), 148L (**e**), 148V (**f**) were demonstrated. The excitation spectra (dashed lines) were detected with the emission wavelength 550 nm and the emission spectra (solid lines) was measured with the excitation wavelength 400 nm in varied pH buffers ranging from 3.5 to 9.0.

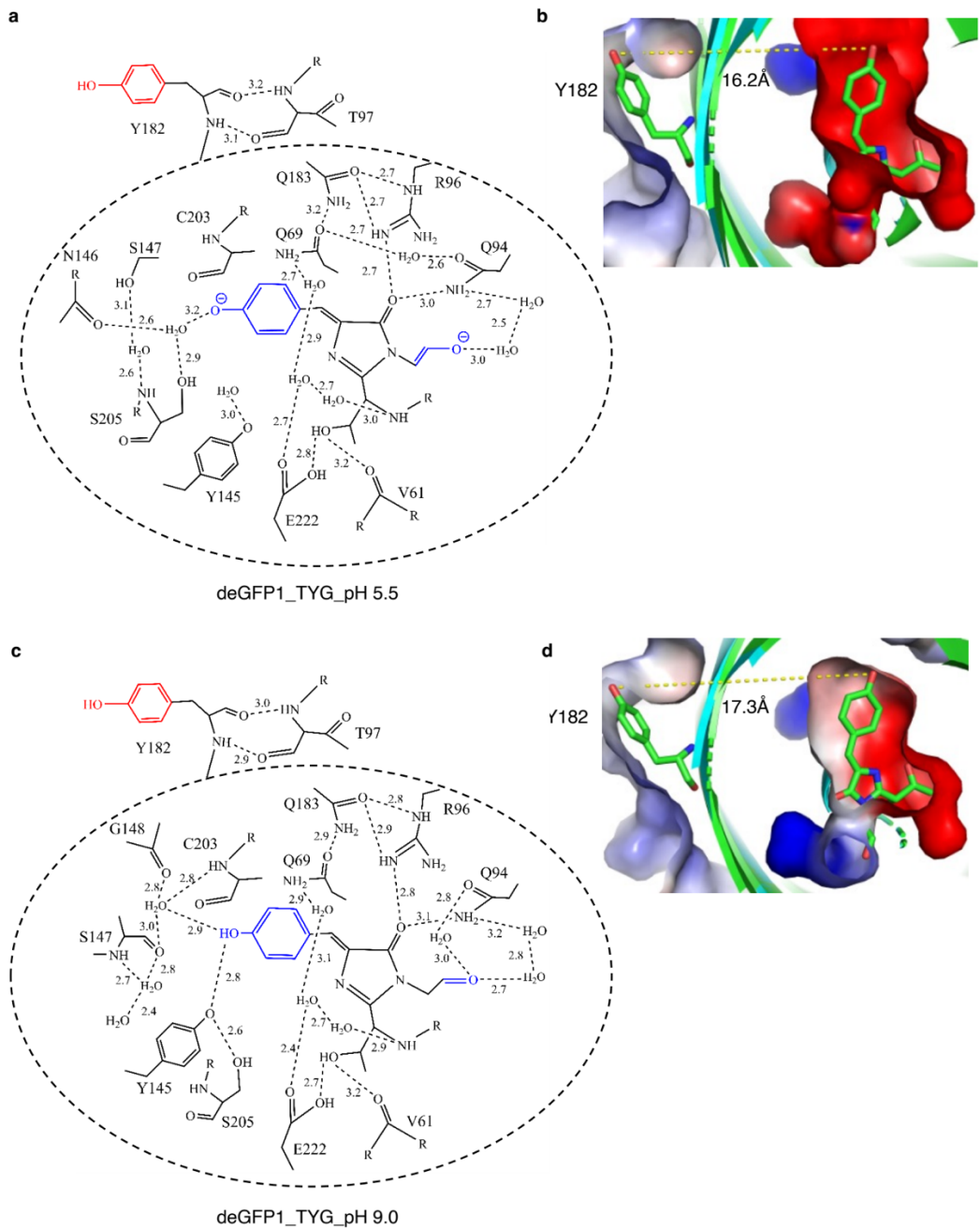

**Supplementary Figure S9. Schematic diagrams depicting the hydrogen-bond network and electrostatic surfaces around the chromophore of deGFP1. a** and **b** The proposed hydrogen-bond network (**a**) and the electrostatic surface (**b**) of the deGFP1 around its chromophore at the crystallized buffer of pH 5.5. **c** and **d** The proposed hydrogen-bond network (**c**) and the electrostatic surface (**d**) at the crystallized buffer of pH 9.0. The long dashed lines represented the distance between the phenolic hydroxyl moieties of Y182 and the chromophore.

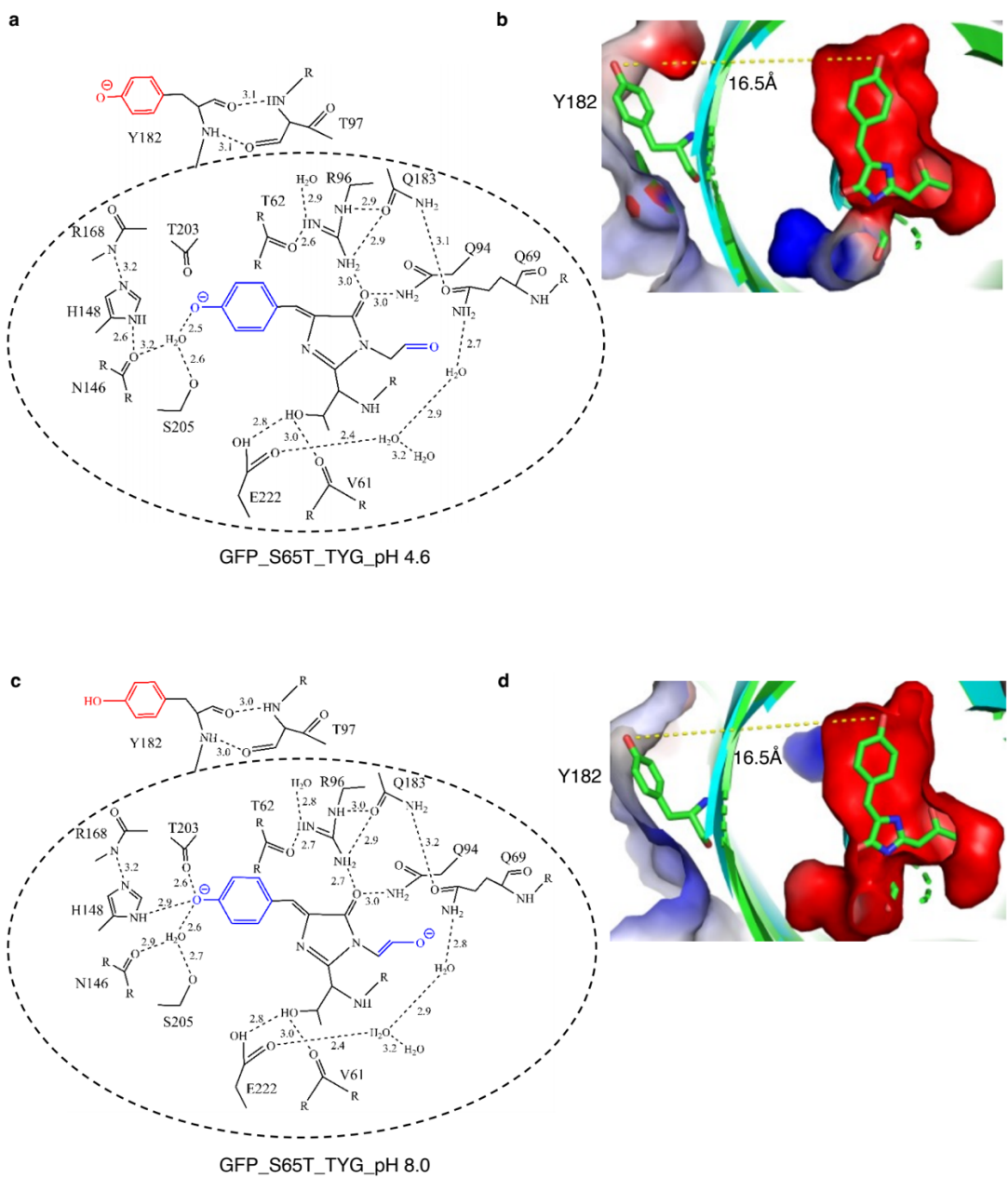

**Supplementary Figure S10. Schematic diagrams illustrating the hydrogen-bond network and electrostatic surfaces around the chromophore of GFP S65T. a** and **b** The proposed hydrogen-bond network (**a**) and the electrostatic surface (**b**) of the GFP S65T around its chromophore at the crystallized buffer of pH 4.6. **c** and **d** The proposed hydrogen-bond network (**c**) and the electrostatic surface (**d**) at the crystallized buffer of pH 8.0. The long-dashed lines represented the distance between the phenolic hydroxyl moieties of Y182 and the chromophore.

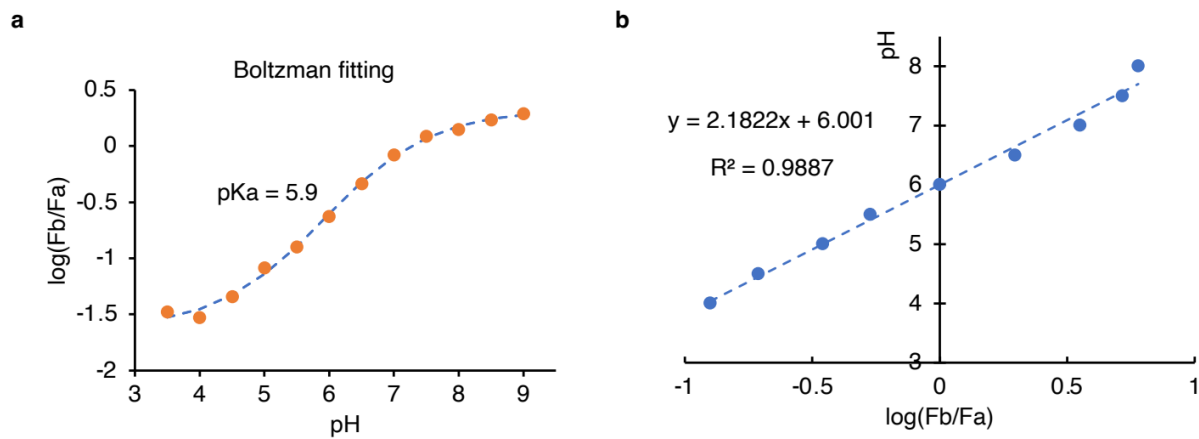

**Supplementary Figure S11. The pKa and pH ultra-sensitivity of SITE-pHorin. a** Boltzmann fitting for pKa with the logarithmic data of the ratio of fluorescent intensities (log (Fb/Fa) in varied pH buffers from 3.5 to 9.0. **b** pH sensitivity plotting and fitting. Data were normalized with value around the pKa of SITE-pHorin. Fb, the fluorescence intensity of the emission wavelength 500-570 nm; Fa, the fluorescence intensity of the emission wavelength 420-480 nm; the Fb and Fa were measured with a single excitation at 405 nm.

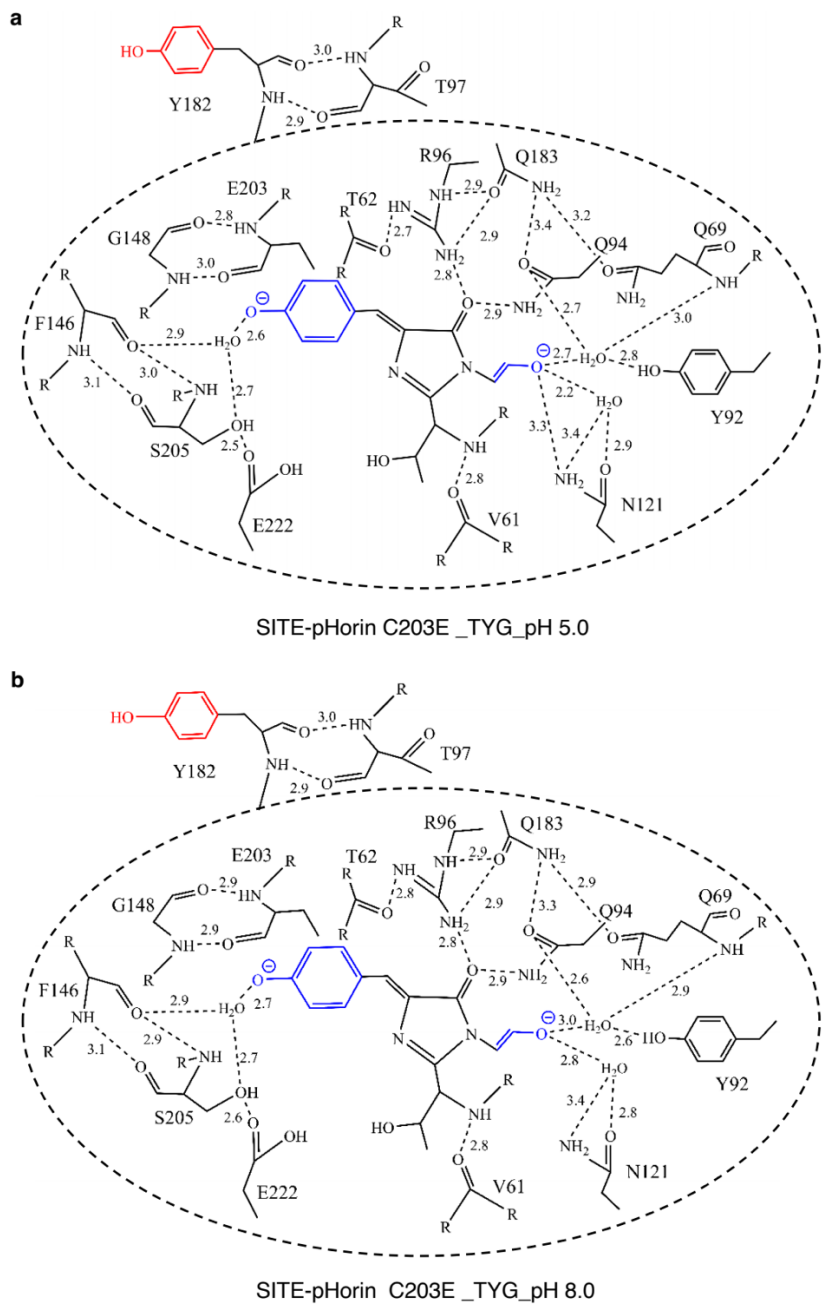

**Supplementary Figure S12. Schematic diagrams depicting hydrogen-bond network around the chromophore of SITE-pHorin mutant (C203E). a** and **b** The proposed hydrogen-bond network of SITE-pHorin mutant (C203E) around its chromophore at the crystallized buffer of pH 5.0 (**a**) and pH 8.0 (**b**), respectively.

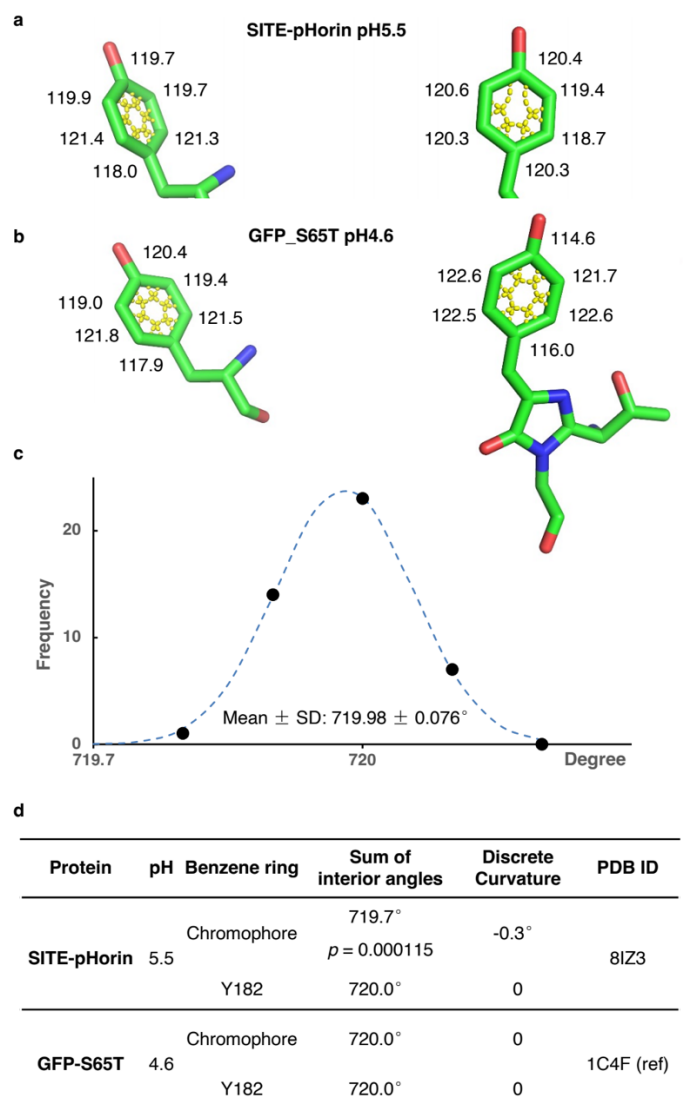

**Supplementary Figure S13. Analysis of benzene rings in chromophore and Y182 of SITE-pHorin and GFP S65T. a** Internal angle degrees of the benzene rings in Y182 and the chromophore in the crystal structure of SITE-pHorin under pH 5.5 conditions were illustrated. **b** Internal angle degrees of the benzene rings in Y182 and chromophore in GFP S65T under pH 4.6 conditions were also displayed. **c** The distribution of the internal angle sums of benzene rings in phenylalanine and tyrosine amino acids in SITE-pHorin were calculated. The results indicated that the sum of these internal angles followed a Gaussian distribution. The fitted mean ± standard deviation of these sums was 719.98 ± 0.076 degree. **d** A summary was tabulated for the internal angle sums of benzene rings of the chromophore and Y182 in the crystal structure of SITE-pHorin at pH 5.5 (PDB: 8IZ3) and GFP S65T at pH 4.6 (PDB: 1C4F) . Interestingly, it was demonstrated that the chromophore of SITE-pHorin at pH 5.5 exhibited a significantly distorted sum of 719.7° from 720.0° ($p = 0.000115$, calculated using the fitted parameters in panel **c**), which resulted in a discrete curvature of -0.3° in the chromophore of SITE-pHorin at pH 5.5.

| Organelle | pH | Methods | Reference |
| --- | --- | --- | --- |
| Cytoplasm | ~7.1 | Fluorescent dye SNARF-1 | (Ramshesh and Lemasters, 2018) |
| | ~7.4 | Enhanced yellow fluorescent protein EYFP | (Llopis et al., 1998) |
| | ~7.5 | Red fluorescent protein mCherryEA | (Rajendran et al., 2018) |
| Nucleus | ~7.1 | Fluorescent dye SNARF-1 | (Ramshesh and Lemasters, 2018) |
| | 7.07~7.80 | Fluorescent dye SNARF-1-AM, SNARF-1-dextran | (Seksek and Bolard, 1996) |
| | ~7.4 | Hoechst-tagged fluorescein | (Nakamura and Tsukiji, 2017) |
| Mitochondria | $8.0 \pm 0.9$ | Fluorescent dye SNARF-1 | (Ramshesh and Lemasters, 2018) |
| Mitochondrial inter membrane space | $6.88 \pm 0.09$ | GPD fused EYFP in human ECV304 cells | (Porcelli et al., 2005) |
| Mitochondrial matrix | ~7.78 | EYFP fused with 12 N-terminal residues cytochrome c oxidase subunit IV and measured in human ECV304 cells | (Porcelli et al., 2005) |
| | $7.98 \pm 0.07$ $7.91 \pm 0.16$ | EYFP fused with cytochrome c oxidase subunit IV targeting signal and measured in Hela and rat cardiomyocytes | (Llopis et al., 1998) |
| | $7.6 \pm 0.45$ | Yellow fluorescent protein (SypHer) with 5-(and 6)-carboxy-SNARF-1 | (Poburko et al., 2011) |
| Golgi | $6.17 \pm 0.02$ | Fluorescent dye $C_6$-NBD-ceramide | (Seksek et al., 1995) |
| | $6.45 \pm 0.03$ | Fluorescein isothiocyanate (FITC) with VTI1B | (Kim et al., 1996); (Schapiro and Grinstein, 2000) |
| Trans-Golgi | 6.4~6.81 | EYFP with trans-Golgi integral membrane protein | (Llopis et al., 1998) |
| | $5.9 \pm 0.07$ | Fluorescent protein RpHluorin2 combined with GalT | (Linders et al., 2022) |
| Medial-Golgi | $6.1 \pm 0.07$ | Fluorescent protein RpHluorin2 combined with MGAT2 | (Linders et al., 2022) |
| ER | $7.07 \pm 0.02$ | Fluorescent dye DTAF-conjugated B-Glyc-KDEL | (Kim et al., 1998) |
| | $7.2 \pm 0.08$ | Fluorescent protein marker ER-RpHluorin2 | (Linders et al., 2022) |
| | 7.1 | Green fluorescent protein pHlurion | (Reifenrath and Boles, 2018) |
| Lysosome | $4.45 \pm 0.10$ | a plasmon Raman pH probe of autophagic HeLa cells | (Li et al., 2019) |
| | 5.0~ 5.4 | pHLARE (pH Lysosomal Activity REporter) of cancer cell | (Webb et al., 2021) |
| | ~4.1 | Cyan fluorescent protein variant mTFP1 | (Chin et al., 2021) |
| | 4.7 ± 0.15 | Fluorescent protein RpHluorin2 combined with LAMP1 | (Linders et al., 2022) |
| Peroxisome | 6.9~7.1 | Fluorescent protein pHluorin-SKL | (Jankowski et al., 2001) |
| | 8.2 ± 0.3 | Fluorescein (5- and 6-) carboxy-SNAFL-2 moiety to peroxisome-targeting sequence (PTS1) | (Dansen et al., 2000) |

**Table S1. The organellar pH measured with various strategies.**

|  | mTurquoise2 S65T | mTurquoise2 W66Y | SITE-pHorin _pH7.0 | SITE-pHorin _pH5.5 | SITE-pHorin C203E_pH5.0 | SITE-pHorin C203E_pH8.0 |
|---|---|---|---|---|---|---|
| **Data collection** | | | | | | |
| Wavelength | 0.979 | 0.979 | 0.979 | 0.979 | 0.979 | 0.979 |
| Space group | P 21 21 21 | P 21 21 21 | P 21 21 21 | P 21 21 21 | P 21 21 21 | P 21 21 21 |
| a, b, c (Å) | 51.224 62.271 70.502 | 51.503 63.357 66.428 | 51.071 62.092 70.658 | 50.925 62.28 70.477 | 51.454 61.759 69.655 | 54.024  62.896 67.590 |
| α, β, γ (°) | 90  90  90 | 90  90  90 | 90  90  90 | 90  90  90 | 90  90  90 | 90  90  90 |
| Resolution (Å) | 50.00-2.00 (2.03-2.00)* | 50.00-2.10 (2.14-2.10) | 50.00-2.30 (2.34-2.30) | 50.00-2.18 (2.26-2.18) | 50.00-1.66 (1.69-1.66) | 50.00-1.57 (1.60-1.57) |
| Completeness (%) | 99.9 (99.4) | 99.9 (99.5) | 99.8 (98.2) | 99.07 (91.26) | 99.9 (99.6) | 99.4 (87.9) |
| $<I>/\delta(I)$ | 11.5 (1.67) | 13.21 (2.05) | 6.71 (1.19) | 10.67 (1.67) | 34.2 (1.89) | 38.0 (5.78) |
| Redundancy | 10.5 (6.0) | 11.9 (8.9) | 10.1 (4.5) | 11.8 (8.3) | 12.7 (11.6) | 12.8 (10.9) |
| $R_{merge}$ | 0.230 (0.678) | 0.124 (0.759) | 0.337(1.058) | 0.243(1.666) | 0.057 (0.809) | 0.083 (0.350) |
| $R_{p.i.m}$ | 0.071 (0.290) | 0.049 (0.267) | 0.106(0.538) | 0.086(0.591) | 0.017 (0.242) | 0.024 (0.108) |
| **Refinement** | | | | | | |
| Resolution (Å) | 35.25-1.99 | 34.27-2.10 | 35.33-2.30 | 31.14-2.18 | 34.85-1.66 | 42.24-1.57 |
| Unique reflections | 15931 | 13253 | 10397 | 12029 | 26832 | 32699 |
| $R_{work}/R_{free}$ | 0.175/ 0.220 | 0.173/ 0.199 | 0.177/ 0.227 | 0.184/ 0.180 | 0.170/ 0.220 | 0.162/ 0.182 |
| No. atoms | | | | | | |
| Protein | 1933 | 1834 | 1924 | 1850 | 1873 | 1865 |
| Ligand/ion | 0 | 0 | 0 | 0 | 0 | 0 |
| Water | 203 | 135 | 131 | 130 | 203 | 219 |
| B-factors | | | | | | |
| Protein | 23.639 | 31.228 | 30.853 | 26.360 | 30.067 | 18.322 |
| Water | 31.605 | 36.844 | 32.603 | 29.840 | 40.187 | 28.941 |
| Ramachandran | | | | | | |
| Favored (%) | 98.3 | 97.8 | 97.0 | 97.76 | 98.0 | 98.0 |
| Allowed (%) | 1.7 | 2.2 | 3.0 | 2.2 | 2.0 | 2.0 |
| Outlier (%) | 0.0 | 0.0 | 0.0 | 0.0 | 0.0 | 0.0 |
| R.m.s.deviations | | | | | | |
| Bond lengths (Å) | 0.003 | 0.011 | 0.003 | 0.013 | 0.012 | 0.014 |
| Bond angles (°) | 1.028 | 1.638 | 0.966 | 1.830 | 1.863 | 1.938 |
| PDB ID | 8IYZ | 8IZ0 | 8IYY | 8IZ3 | 8IZ1 | 8IZ2 |

**Table S2. Summary of data collection and refinement statistics.** The symbol * represent the highest resolution shell in parenthesis.

| Marker | Gene ID | Gene fragment for targeting (bp) | Location |
|---|---|---|---|
| hH2B | NM_021058.4 | 1-378 | Nucleus |
| g-subunit | NM_001244137.1 | 1-309 | Mito cristae |
| mFech | M61215.1 | 1-1260 | Mito IMS |
| hST6GAL1 | NM_003032.3 | 1-1218 | Trans-Golgi |
| hMGAT2 | NM_002408.4 | 1-267 | Medial-Golgi |
| hGP73 | NM_016548.4 | 1-102 | Cis-Golgi |
| mCalreticulin | NM_007591.3 | 1-1248 | ER |
| mDnase2B | NM_019957.5 | 1-1062 | Lysosome |
| rTFR1 | NM_022712.1 | 1-2283 | Endosome |

**Table S3. Information about the gene fragments for organelle targeting.** The gene ID was found from the Genebank database. The lengths of gene fragments for organelle targeting fragment were used for molecular clone. The peptides were fused with the N-terminal of SITE-pHorin.

| Protein | pH | Y182 | Chromophore | Interaction energy (Hartree) | |
|---|---|---|---|---|---|
| | | Charge | | Base set | |
| | | | | 6-31G** | Aug-cc-pVDZ |
| SITE-pHorin | 5.5 | -1 | 0 | -0.3345 | -0.3653 |
| | 7.0 | 0 | -1 | -0.0725 | NA |
| GFP-S65T | 4.6 | -1 | -1 | -0.3184 | NA |
| | 8.0 | 0 | -2 | -0.0091 | NA |

**Table S4. The interaction energy between chromophore and Y182.** Calculations were performed using B3LYP/6-31G** or the basis set Aug-cc-pVDZ as computational settings with the software NWChem in quantum chemistry. NA, not available due to converge problem.